\documentclass[11pt,dvips,twoside,letterpaper]{article}
\usepackage{times}
\usepackage{fancyhdr}
\usepackage{geometry}
\usepackage{graphicx}

\usepackage{amsmath,amssymb}
\usepackage{bm}

\begin{document}
\title{
~\\[-1.2in]
{\normalsize\noindent
\begin{picture}(0,0)(208,-3.15)
\begin{tabular}{l p{1.1in} r}
In: Focus on Boson Research  & & ISBN 1-59454-520-0 \\
Editor: A. V. Ling, pp. 1-60  & & \copyright 2005 Nova Science Publishers, Inc.\\
\end{tabular}
\end{picture}}
{\begin{flushleft} ~\\[0.63in]{\normalsize\bfseries\textit{Chapter~1}} \end{flushleft} ~\\[0.13in] 
\bfseries\scshape Bosons after Symmetry Breaking \\
in Quantum Field Theory}}
\author{
\bfseries\itshape Takehisa Fujita\thanks{E-mail address: fffujita@phys.cst.nihon-u.ac.jp}, Makoto Hiramoto\thanks{E-mail address: hiramoto@th.phys.titech.ac.jp}, and Hidenori Takahashi\thanks{E-mail address: htaka@phys.ge.cst.nihon-u.ac.jp}\\
Department of Physics, Faculty of Science and Technology,\\
Nihon University, Tokyo, Japan}
\date{}
\maketitle
\thispagestyle{empty}
\setcounter{page}{1}

\pagestyle{fancy}
\fancyhead{}
\fancyhead[EC]{Takehisa Fujita, Makoto Hiramoto and Hidenori Takahashi}
\fancyhead[EL,OR]{\thepage}
\fancyhead[OC]{Bosons after Symmetry Breaking 
in Quantum Field Theory}
\fancyfoot{}
\renewcommand\headrulewidth{0.5pt}
\addtolength{\headheight}{2pt} 
\headsep=9pt

%
%
%

\begin{abstract} 
We present a unified description of the spontaneous symmetry breaking 
and its associated bosons in fermion field theory. There is no Goldstone 
boson in the fermion field theory models of Nambu-Jona-Lasinio, Thirring 
and QCD$_2$ after the chiral symmetry is spontaneously broken in the new vacuum. 
The defect of the Goldstone theorem is clarified, and the "massless boson" 
predicted by the theorem is virtual and corresponds to just a $free$ massless fermion and 
antifermion pair. Further, we discuss the exact spectrum of the Thirring model 
by the Bethe ansatz solutions, and the analytical expressions of all the physical 
observables enable us to understand the essence of the spontaneous symmetry breaking 
in depth. Also, we examine the boson spectrum in QCD$_2$, and show that bosons 
always have a finite mass for $SU(N_c)$ colors. The problem of the light cone 
prescription in QCD$_2$ is discussed, 
and it is shown that the trivial light cone vacuum is responsible for the wrong prediction of 
the boson mass. 
\end{abstract}

%
%
%
%
%
%
%
%
%
%
%
%
%
%
%
%
%
%
%
%

\section{\bf Introduction}

Most of the field theory models should possess some kind of symmetries. 
Apart from the basic symmetry like the Lorentz invariance, there are 
symmetries which play a fundamental role in determining the structure 
of the vacuum state. The chiral symmetry is one of the most popular ones 
in quantum field theory of fermions. 

The physics of the chiral symmetry and its spontaneous symmetry breaking 
in fermion field theory models has been discussed quite extensively since the invent 
of the Goldstone theorem \cite{q1,q2}. 
In particular, the current current interaction model of Nambu and Jona-Lasinio (NJL) 
has been studied by many people since it is believed that the NJL model 
can present a good example of exhibiting the Goldstone boson after 
the chiral symmetry is spontaneously broken \cite{q7}.  

However, recent careful studies clarify that there appears no massless 
boson in the NJL model \cite{q91,q9}. Further, it is shown that the Goldstone theorem 
cannot be applied to the fermion field theory models due to a serious defect 
in the course of proving the theorem \cite{hist}. That is, the existence 
of a massless boson that should be proved as the result of the Goldstone equation 
has to be assumed as the initial input of the equation, and this is of course no proof at all. 

In the NJL model, the Lagrangian density possesses the chiral symmetry, but 
the vacuum state prefers the chiral symmetry broken state since it 
has the lower energy than the one with preserving its symmetry.  
In this procedure, the chiral current is conserved and therefore 
the symmetry breaking is considered to be spontaneous due to  
the definition of the $spontaneous$ symmetry breaking in proving the Goldstone theorem. 
However, if one calculates the boson mass properly, then there appears 
no massless boson in the NJL model. In this case, a question may arise 
as to why people obtained the massless boson in the NJL model. 
Not only Nambu and Jona-Lasinio but also quite a few physicists found 
a massless boson in their boson mass calculation \cite{q07}. Surprisingly, 
the reason why they found a massless boson is simple. 
They calculated the boson mass by summing up one loop Feynman 
diagrams, but their calculation is based on the perturbative vacuum 
state. However, after the spontaneous symmetry breaking, one finds 
the new vacuum which has the lower energy than the perturbative vacuum state. 
Therefore, the physical vacuum state is of course the new vacuum that breaks 
the chiral symmetry, and thus if one wishes to calculate any physical observables 
in field theory, then one must employ the formulation which is based on 
the physical vacuum state. In fact, the quantum field theory is constructed 
on the physical vacuum state, and therefore it is essential to make the formulation 
which is based on the physical vacuum state. Those calculations which start from 
the perturbative vacuum state should find a unphysical boson. Indeed, this unphysical 
massless boson is just what Nambu and all other people obtained in their calculations. 

In this chapter, we review the spontaneous symmetry breaking and 
its appearance of bosons associated with the symmetry breaking phenomena. 
The symmetry breaking of the vacuum occurs in the boson field theory as well 
as in the fermion field theory models. In the boson field theory, the symmetry 
breaking is clear in four dimensions, but the symmetry should not be 
broken in two dimensions since there should not exist a physical massless 
boson due to the infra-red divergence of its propagator. In the fermion field theory, 
the symmetry is broken in the vacuum in two and four dimensions. This is 
quite simple because there is no Goldstone boson and therefore the two 
dimensional field theory is not special any more. 

The picture of the chiral symmetry breaking in the fermion field theory can be drawn 
in the following way.  One starts from the perturbative vacuum which is the same 
as the free vacuum state of fermions. This preserves the chiral symmetry. 
When one takes the interactions into account, then the distributions of the 
negative energy particles in the vacuum state change and the vacuum energy 
becomes lower than that of the free vacuum state. The momentum distributions of 
the negative energy particles become different for right and left mover fermions 
so that the chiral symmetry is broken in the new vacuum. In addition, 
the distribution of the negative energy particles in the new vacuum is rearranged 
such that the finite gap is seen in the excitation spectrum. This is essentially all 
that happens to the symmetry breaking phenomena in the fermion field theory models 
like NJL, Thirring model, QED$_2$ and QCD$_2$ with massless fermions. 
On this vacuum state, one may find bosons or there may be no boson, and this depends 
on the interactions between fermions and antifermions. For the current current interaction 
models like the NJL or the massless Thirring model, one cannot make any bosons since 
the interaction is a $\delta-$function type potential. 
On the other hand, the gauge field theory models of QED$_2$ and QCD$_2$ find 
massive bosons since the interaction is a confining potential in two dimensions. 
At this point, it is probably fair to mention that, if one finds the true vacuum 
state in quantum field theory model, then it means that one could solve this field 
theory model exactly. However, the NJL model is not exactly solvable, and we believe 
the solution of the vacuum state constructed by Bogoliubov transformation method 
is not exact. However, the new vacuum state has the lower energy than the perturbative 
vacuum state, and the qualitatively right picture of the symmetry breaking phenomena 
should be obtained from this approximate vacuum state.  

For the symmetry broken vacuum states, the chiral condensate value is finite 
in all of the massless fermion field theory models. This is a representation 
of the vacuum structure, and it is quite natural that the symmetry broken vacuum 
state has a complicated vacuum structure with a finite condensate value. 
However, QED$_2$ and QCD$_2$ with massive fermions have no chiral symmetry, and 
therefore one cannot discuss about the symmetry broken vacuum. Nevertheless, 
the vacuum state has a finite condensate value, and it keeps the complicated 
vacuum structure. In these field theory models of QED$_2$ and QCD$_2$, the fermion 
mass term seems to play a role of just like a perturbative interaction term. 
The boson mass increases linearly as the function of the fermion mass $m_0$. 
The behavior of the chiral condensate is somewhat similar to the boson mass, and 
it decreases as the function of the fermion mass $m_0$ and the condensate value 
goes to zero at the very large fermion mass where the system becomes nonrelativistic. 
In this respect, it is still not very clear how the chiral condensate value 
can be related to the symmetry breaking phenomena.  

This chapter is organized in the following way. In the second section, 
we review the Goldstone theorem and discuss its problem related to the fermion field 
theory models. This theorem is originally meant for the boson field 
theory models, but it was believed that the theorem should 
be valid for the fermion field theory models as well. However, one 
can easily notice the basic defect of the theorem, that is, the existence 
of the massless boson that should be proved as a final goal has 
to be assumed in the initial equation.  
Here, we present a proof why the Goldstone theorem is justified for the boson field case
while it does not hold for the fermion field case.

In section 3, we treat the spontaneous symmetry breaking in boson field theory models. 
This is well established in four dimensions, and we present it only for the clarification of 
the essence of the physics of the spontaneous symmetry breaking phenomena and 
the appearance of the massless boson associated with the symmetry breaking. 
However, the symmetry breaking and the spectrum of the boson field theory 
in two dimensions is not so clear as in four dimensions, 
and we discuss problems behind the two dimensional boson fields. 

In section 4, we discuss the non-appearance of a massless boson 
after the spontaneous symmetry breaking in fermion field theory models. 
Since the spontaneous symmetry breaking has a long history, the non-appearance 
of the Goldstone boson is indeed a surprising result. 
But it was due to the lack of the deep understanding of the vacuum structure, 
and, in a sense, it should have been very difficult to clearly 
realize the importance of the change in the vacuum structure which arises 
from the symmetry breaking. Here, we present the symmetry breaking and 
its boson associated with the symmetry breaking in the NJL and the Thirring models. 
We first give intuitive discussions why the chiral symmetry is broken in the vacuum of the 
NJL and the Thirring models, and clarify why there should not appear any 
massless boson. Further, we carry out more elaborate calculations of the symmetry 
breaking in these fermion field theory models based on 
the Bogoliubov transformation method. In this calculation, one sees that 
there should be a massive boson depending on the strength of the coupling constant, 
but the boson is not a consequence of the symmetry breaking, but it is due to 
the result of the approximate scheme of the Bogoliubov transformation. It is most 
probable that there should be no boson in the NJL and the Thirring models. 

In section 5, we present the Bethe ansatz solutions 
of the massless Thirring model. This is quite important to understand 
the structure of the new vacuum and its change of the negative energy 
particle distribution in the vacuum state. One clearly sees that the new 
vacuum state that breaks the chiral symmetry has the lower energy than the 
symmetric vacuum state. In this case, the momentum distribution of 
the negative energy state changes drastically. This means that the symmetry 
is spontaneously broken even though the Thirring model is a two dimensional 
field theory model. However, there is neither a massless boson nor 
a massive boson. There is only a finite gap for the excitation spectrum. 
In this respect, one can learn a lot about the symmetry breaking and 
its boson associated with the spontaneous symmetry breaking.   
The non-existence of a massless boson is very reasonable since there should 
not exist any massless boson in two dimensions. Here, we also show 
that the bosonization procedure which is commonly used for the massless 
Thirring model has a serious defect in that one cannot find 
the corresponding degree of freedom for the zero mode. Therefore, 
the massless Thirring model is not bosonized properly, and therefore 
it is indeed consistent with the finite gap in the spectrum. 

In section 6, we discuss the Schwinger model, and one knows that it is 
well understood. There is no new thing added to this section. 
However, we believe that it should be important to understand the origin 
of the chiral symmetry breaking. In the Schwinger model, the chiral symmetry 
is broken, but it is not spontaneous since the chiral current is 
not conserved any more due to the anomaly term after the regularization 
of the vacuum state. The anomaly term arises from the conflict between 
the gauge invariance and the chiral current conservation. However, 
it is interesting to note that the chiral condensate is a smooth function 
of the fermion mass, and it is not very clear yet whether the chiral 
condensate is a consequence of the chiral symmetry breaking or not. 

In section 7, we present the recent results of the numerical calculations 
of SU($N_c$) QCD in two dimensions in terms of the Bogoliubov transformation method. 
The calculations are carried out up to 
very large values of the $N_c$ color degree, and it is shown that the SU(50) 
calculation is already quite similar to the result with the $N_c \rightarrow \infty $.  
However, it turns out that the light cone calculation cannot reproduce 
neither the right boson spectrum nor the right condensate values. 
This must be due to the fact that the light cone vacuum is trivial even 
though the real vacuum has a complicated structure with a finite condensate 
value. In fact, 't Hooft calculation is not an exception and gives wrong results 
of the spectrum since he employed the light cone vacuum, even though the large $N_c$ 
expansion itself is a right and good scheme for SU($N_c$) QCD$_2$. 
Therefore, if one wishes to find the correct spectrum of the field theory model, 
then one has to start from the right vacuum state as a minimum condition. 
Here, the calculated results of the spectrum and the condensate values 
by the Bogoliubov transformation method indicate that the symmetry is spontaneously 
broken in the vacuum state, and there is no massless boson.  This is again consistent 
with the fact that there should be no physical massless boson in two dimensions.

In section 8, we summarize what we have clarified in the spontaneous symmetry breaking and 
its boson associated with the symmetry breaking phenomena. Some comments on the Heisenberg 
XXZ model and the lattice field theory are included.


\section{ Goldstone Theorem and its Applicability }

The Goldstone theorem has played a central role 
for understanding the symmetry breaking and its massless 
boson after the spontaneous symmetry breaking. 
When the Lagrangian density has some continuous symmetry which can 
be represented by the unitary operator $U(\alpha)$, there is a conserved 
current associated with the symmetry 
$$ \partial_\mu j^\mu =0 . \eqno{(2.1)} $$
In this case, there is a conserved charge $ Q $ which is defined as 
$$ Q = \int j^0 (x) d^3x . \eqno{(2.2)} $$ 
The Hamiltonian of this system $H$ is invariant under the unitary transformation 
$U(\alpha)$, 
$$  U(\alpha)H U(\alpha)^{-1} =H .  \eqno{(2.3a)}  $$
Writing the $U(\alpha)$ explicitly as $U(\alpha)=e^{i\alpha Q} $, we obtain 
$$ QH=HQ .  \eqno{(2.3b)} $$
Now, the vacuum state can break the symmetry, and we define the symmetric vacuum 
$|0\rangle $ and  symmetry broken vacuum $|\Omega \rangle $, respectively, which 
satisfy the following equations, 
$$ U(\alpha) |0\rangle =|0\rangle  \eqno{(2.4a)}  $$
$$ U(\alpha) |\Omega \rangle \not= |\Omega \rangle . \eqno{(2.4b)}  $$
These equations can be written in terms of the charge operator $Q$ as
$$ Q |0\rangle =0  \eqno{(2.5a)}  $$
$$ Q |\Omega \rangle \not= 0 . \eqno{(2.5b)}  $$
In the Goldstone theorem, it is assumed that the current conservation arising from the symmetry 
should hold after the symmetry is broken in the vacuum state. Therefore,  the commutation 
relation between the charge and some boson field operator $\phi (x)$ 
is time independent, and 
we write it as 
$$  [Q(t), \phi (x)] = \hat{ C}  \eqno{(2.6)}  $$
where $\hat{ C} $ is some operator that is described by the field operators. 
This is of course an identity equation, but the choice of the field  $\phi (x)$ 
itself is not at all trivial. It should also be important to note that eq.(2.6) is 
derived independently from the Hamiltonian of field theory models. 

Now, we take the expectation value of eq. (2.6) with the vacuum state, and there 
the information of the field theory model should be put in eq.(2.6) through the vacuum state. 
First, we employ the symmetric vacuum $|0\rangle $, 
$$ \langle 0 \mid [Q(t), \phi (x)] \mid 0 \rangle =
\langle 0 \mid \hat{C} \mid 0 \rangle . \eqno{(2.7a)} $$
In this case, the left hand side vanishes. Therefore, the right hand side must also vanish, 
and eq. (2.6) gives just the identity equation as expected. 

Next, we take the expectation value of eq.(2.6) with the symmetry broken vacuum 
$|\Omega \rangle $, 
$$ \langle \Omega \mid [Q(t), \phi (x)] \mid \Omega \rangle =
\langle \Omega \mid \hat{C} \mid \Omega \rangle \not= 0 . \eqno{(2.7b)} $$
If the right hand side is non-zero, then the vacuum of the system 
has the symmetry broken state since the left hand side survives only 
when the operator $Q$ satisfies eq.(2.5b). 

The Goldstone theorem  starts from  the vacuum expectation 
value of the commutation relation eq.(2.7b) with the symmetry broken vacuum $|\Omega \rangle $, 
and the boson field $\phi$ eventually corresponds to a massless boson. 

Further, we assume that the field $\phi$ and the current density $j^0(x)$ satisfies the following translational 
property 
$$ \phi (x) = e^{ipx} \phi (0) e^{-ipx}  \eqno{(2.8a)} $$ 
$$ j^0(x) = e^{ipx} j^0(0) e^{-ipx} . \eqno{(2.8b)} $$ 
Now, we insert a complete set of intermediate bosonic states $|n\rangle$ in eq.(2.7b). 
Since it should be excited by the charge operator $Q$, this state $|n\rangle$ 
should have the same momentum as the vacuum state, and this means that the momentum 
of the bosonic state $|n\rangle$  is zero.  

Thus, we obtain from eq.(2.7b),
$$ \sum_n {(2\pi)}^3 \delta (\bm{p}_n )\biggl[  \langle \Omega |j^0(0)| n \rangle 
\langle n | \phi (0) | \Omega \rangle e^{-iE_nt} 
 - \langle \Omega |\phi (0)| n \rangle \langle 
n | j^0(0) | \Omega \rangle e^{iE_nt} \biggr] 
 \neq 0  . 
\eqno{(2.9)} $$
The right hand side is non-zero and is also time-independent. 
However, in the left hand side, the positive and negative energy terms cannot  
cancel with each other as long as the energy $E_n$ is non-zero. Therefore, the time dependence 
of eq.(2.9) in the left hand side can be taken away only when the following condition is satisfied, 
$$ E_n =0 \quad {\rm for} \quad \bm{p}_n =0 . \eqno{(2.10)} $$
From this constraint, one learns that if this bosonic state is an isolated system, 
then this should correspond to a massless boson.  
Thus, there should appear a massless boson after the spontaneous symmetry breaking.  
However, there is a serious difference 
between the boson field and fermion field theory models, and we show 
that the proof of the Goldstone theorem cannot be applied to 
the fermion field theory models since the existence of the boson field has to be 
assumed while the boson field itself is, however, the one that must be proved 
as a result. 

The difficult part in eq.(2.9) is to find the boson field operator  $\phi$, and 
if one can find it properly, then one can obtain some physical information 
from the identity equation. It should be noted that, normally, 
one cannot get any important information from the identity equation 
since it is not directly related to the dynamics of the field theory model.  

At this point, we should comment on a possible degeneracy of the symmetry broken 
vacuum state associated with the charge operator $Q$ in the total Hamiltonian system 
since there is a belief that the symmetry broken vacuum  may have infinite degenerate 
states. First, we define the symmetry broken vacuum energy 
which is the eigenstate of the Hamiltonian $H$, 
$$ H|\Omega \rangle =E_\Omega |\Omega \rangle . \eqno{(2.11)} $$
Now, if we define a new state $|\varphi_n \rangle $ by 
$$ |\varphi_n \rangle \equiv {\cal N} Q^n |\Omega \rangle \eqno{(2.12)} $$
with ${\cal N}$ a normalization constant, then the state $|\varphi_n \rangle  $ 
has the same vacuum energy $E_\Omega $ because of eq.(2.3b), 
$$ H|\varphi_n \rangle = {\cal N}HQ^n|\Omega \rangle  = {\cal N}Q^nH|\Omega \rangle =E_\Omega 
|\varphi_n \rangle . \eqno{(2.13)} $$
This equation seems to indicate that the vacuum state has $n-$degeneracy. 
However, $Q$ has the same eigenstate as $H$ due to eq.(2.3b), and therefore we write 
with its eigenvalue $q$
$$ Q |\Omega \rangle = q |\Omega \rangle .  \eqno{(2.14)} $$
Thus, we obtain
$$ |\varphi_n \rangle ={\cal N}{q}^n |\Omega \rangle = |\Omega \rangle \eqno{(2.15)} $$ 
since we can choose the normalization constant $\cal N$ as 
${\cal N}={q}^{-n} $. Therefore, the state $ |\varphi_n \rangle $ 
is nothing but the vacuum state $|\Omega \rangle$ itself, and there is no 
degeneracy because the charge operator $Q$ cannot change the vacuum structure. 
This degeneracy is spurious, but the degeneracy of the potential 
vacuum in the double well potential problem in the boson field theory model 
is real, and this will be discussed in section 3.


\subsection{ Boson Field Theory Model }

The Goldstone theorem is proved by employing eqs.(2.7) and (2.8). 
In the boson field theory models, the boson field $\phi $ exists 
from the beginning. This is of course a trivial thing in the boson 
field theory models. This boson field $\phi $ in eq.(2.7) corresponds 
eventually to the Goldstone boson. Here, the most important point 
is that boson is characterized by its mass, and therefore the determination 
of the boson mass is the only concern for the boson field theory models.  
Also, eq.(2.8) has no problem since the boson should exist as an elementary 
boson state and therefore it satisfies the translational invariance. 

From eqs.(2.9), one obtains the constraint on the bosonic state as given by 
eq.(2.10). This constraint does not necessarily mean that this bosonic state should have 
the dispersion relation of a massless boson. But if this system is an isolated one, 
then this is just the dispersion relation of a massless boson, and therefore 
there should be a massless boson, and this is just the Goldstone boson.    


\subsection{ Fermion Field Theory Model }

Now, we discuss the fermion field theory models \cite{hist}. It is important to note 
that, in the fermion field theory models, bosons must be constructed 
by the fermions and antifermions as their bound states. There is no 
elementary boson field in this field theory model itself. 

In this case, we should ask ourselves what is the boson field $\phi $ 
in eq.(2.7) in the fermion field theory models. 
This boson field $\phi $ should eventually correspond to a massless 
boson if at all exists in the fermion field theory models. But who shows 
that there are any bound states of the fermions and antifermions in this 
field theory models ? This should involve dynamics and it should be very hard 
to solve any of the fermion field theory models until one finds bound states. 

Now, in the proof of the Goldstone theorem, the existence of the boson field $\phi $ 
is assumed, and this is just the one that should be proved as a final goal. 
Thus, it is clear that eq.(2.7) cannot be applied to the fermion field theory 
models. Eq.(2.7) can give one information which is the dispersion 
relation, but one cannot take out the information on the existence 
of the bound state between fermions and antifermions.  Further, if one wishes to 
evaluate the commutation relation between the conserved charge and the boson field $\phi$ 
in eq.(2.7), then one has to be able to describe the boson field $\phi$ 
in terms of the fermion field operators. This is quite clear since the charge 
$Q_5(t) = \int j_5^0 (x) d^3x$ is written in terms of the fermion field operators, 
and therefore one should have the expression of the boson field $\phi$ 
by the fermion fields. 
$$ \phi = F[\bar{\psi}, \psi]  . \eqno{(2.16)} $$
The functional dependence of $ F[\bar{\psi}, \psi]$ should be determined 
by solving the dynamics, and it should be extremely difficult to find 
the functional dependence of $ F[\bar{\psi}, \psi]$. In fact, it is practically 
impossible to find the functional dependence of $ F[\bar{\psi}, \psi]$ unless 
the field theory model is exactly solvable. In two dimensional field theory 
models, there is one example which is solved exactly, and that is the Schwinger 
model as we treat it in section 6. In this case, one can describe the boson 
field in terms of the fermion field operators. 

Here, we show a common mistake which is often found in the textbook to describe how 
the Goldstone theorem holds in fermion field theory model. One says 
that one may take the following $\phi$ in the case of the chiral 
symmetry breaking
$$  \phi(x) = \bar \psi(x) \gamma_5 \psi(x) . \eqno{ (2.17)}   $$
In this case, one can easily prove that this $\phi$ satisfies eq.(2.6). 
In fact, one obtains for eq. (2.6)
$$  \left[Q_5(t), \bar \psi(x) \gamma_5 \psi(x) \right] = 
2 \bar \psi(x)  \psi(x)  \eqno{(2.18)}  $$
where $Q_5$ denotes the chiral charge which is a conserved quantity 
for the chiral symmetry preserving system. It should be quite important 
to note that eq.(2.18) is derived independently from the shape of the interaction 
Lagrangian density. The only condition is that the interaction term should be 
invariant under the chiral symmetry. 
Therefore, eq.(2.18) does not carry any information about the dynamics 
of the fermion and anti-fermion system, and therefore there is no chance 
that one obtains any information about the bound state of fermion and 
ant-fermion from eq.(2.18).  

If we take the expectation value of eq.(2.18) with the symmetry broken vacuum 
state, then we obtain
$$ \sum_n {(2\pi)}^3 \delta (\bm{p}_n )\biggl[  \langle \Omega |j_5^0| n \rangle 
\langle n |\bar \psi \gamma_5 \psi  | \Omega \rangle e^{-iE_nt} 
 - \langle \Omega |\bar \psi \gamma_5 \psi | n \rangle \langle 
n | j_5^0 | \Omega \rangle e^{iE_nt} \biggr]  \neq 0   \eqno{(2.19)} $$
where $|n \rangle $ denotes the complete set of the fermion number zero states 
of the field theory model one considers. 
Therefore, bosonic states as well as the massless free fermion and 
antifermion states should be included in the intermediate states. 
Eq.(2.19) is just the same equation as the boson case, 
and therefore, it gives an impression that the Goldstone theorem is meaningful 
for the fermion field theory models as well. 

However, one easily notices that this $\phi$ has nothing to do with 
any bound state of fermions and antifermions since, as we mention above, 
eq. (2.19) does not contain any information on the interaction term. 
Namely, we obtain from eq.(2.19),
$$ E_n =0 \quad {\rm for} \quad \bm{p}_n =0  \eqno{(2.20)} $$
where 
$$ E_n=E_f+E_{\bar f} \eqno{(2.21a)} $$
$$ \bm{p}_n=\bm{p}_f+\bm{p}_{\bar f}  \eqno{(2.21b)} $$
where $\bm{p}_f$ ($\bm{p}_{\bar f}$) and $E_f$ ($E_{\bar f}$) denote 
the momentum and energy of the fermion (anti-fermion), respectively. 
For the free massless fermion and anti-fermion pair, eq.(2.20) is indeed satisfied. 
This energy dispersion looks like 
a massless boson, but of course it has nothing to do with the massless boson. 

Obviously, the existence of the boson field $\phi$ can be confirmed only after 
the whole dynamics of this field theory model is completely solved. 
As mentioned above, eq. (2.19) does not contain any information on the interaction 
term of the Lagrangian density, and therefore it is natural that eq. (2.19) cannot 
prove  the existence of the bosonic states in the corresponding field theory model. 
Further, even if one could solve the dynamics properly, it would not mean that 
the  $\phi$ can be expressed in terms of fermion field operators.

\section{\bf Goldstone Boson in Boson Field Theory}

The physics of the spontaneous symmetry breaking started from the boson field theory 
models, and Goldstone discovered that there should appear a massless boson 
when the symmetry is spontaneously broken in the vacuum state. The basic point 
in this theorem is that the vacuum state always prefers the lowest energy state 
of the total Hamiltonian and therefore when the minimum energy state of 
the interaction field energy is located at the point which breaks the symmetry 
of the Lagrangian density, then one should find the vacuum state 
which breaks this symmetry. 
The interesting discovery of Goldstone is that there should appear a massless boson 
when one adds the kinetic energy terms of the boson field to the interaction field 
energy term. This is quite similar to the situation where the degenerate states in quantum 
mechanics are split into several states due to the perturbative interaction. Here, 
the role of the perturbative interaction is played by the boson's kinetic 
energy terms, and this is quite important to realize since the degeneracy is 
resolved by the kinetic energy term of the boson field, and thus this indeed leads to 
a massless boson in the Goldstone theorem. In this respect, one says that the degrees 
of freedom of the degeneracy of the symmetry become the Goldstone boson since 
the interaction that breaks the degeneracy is the boson's kinetic energy term. 

\subsection{Symmetry Breaking in Four Dimensional Boson Fields} 

Now the discussion of the spontaneous symmetry breaking in boson 
field theory in four dimensions can be found in any field theory text books, and therefore 
we only sketch the simple picture why the massless boson appears 
in the spontaneous symmetry breaking. 

The Hamiltonian density for complex boson fields can be written as 
$$ {\cal H} = {1\over 2} (\bm{p}{\phi}^{\dagger})(\bm{p}{\phi})+
U\left( |{\phi}| \right) . 
\eqno{(3.1)} $$
This has a $U(1)$ symmetry, and when one takes the potential as
$$ U\left( |{\phi}| \right) = U_0 \left(  |{\phi}|^2 
-\lambda^2 \right)^2 \eqno{(3.2)} $$
then, the minimum of the potential $ U\left(  |{\phi}| \right)$ 
can be found at $|\phi|=\lambda$. But one must notice that this is 
a minimum of the potential, but not the minimum of the total energy. 

The minimum of the total energy must be found together with the kinetic 
energy term. 
When one rewrites the complex field as 
$$ \phi = (\lambda + \rho) e^{i{\xi\over{\lambda}}} \eqno{(3.3)} $$
then, one can rewrite eq.(3.1) as
$$ {\cal H} = {1\over 2}\bigl[ (\bm{p}{\xi})(\bm{p}{\xi})
+(\bm{p}{\rho})(\bm{p}{\rho}) \bigr]
+U\left( |\lambda + \rho| \right) +... \eqno{(3.4)} $$
Here, one finds the massless boson $\xi$ which is associated with the degeneracy  
of the vacuum energy. 
The important point is that this infinite degeneracy of the potential vacuum 
is converted into the massless boson degrees of freedom when 
the degeneracy of the potential vacuum is resolved by the kinetic energy term. 

Also, it should be noted that the massless boson appears at the time when 
the new vacuum is determined. Namely, the massless boson and the new 
vacuum creation after the spontaneous symmetry breaking should occur 
at the same time because the appearance of the Goldstone boson is 
the consequence of the symmetry breaking of the vacuum state.  
Eq.(3.4) shows that the excitation spectrum of the boson system with respect 
to the field $\rho$ has nothing to do with the symmetry breaking.

\subsection{Symmetry Breaking in Two Dimensional Boson Fields} 

The spontaneous symmetry breaking should not occur in two dimensional 
field theory models due to Coleman's theorem \cite{q3}. 
However, as discussed in section 2, there appears no massless boson 
after the symmetry breaking in fermion field theory models, 
and therefore the symmetry can be spontaneously broken 
in two dimensional field theory models of fermions. Indeed, the massless 
Thirring model and two dimensional QCD breaks the chiral symmetry 
in the vacuum state, and there is no massless boson in these models. 

Now, the boson field theory models in two dimensions cannot break 
the symmetry spontaneously since there should appear a massless boson 
associated with the symmetry breaking. In this case, however, it is 
interesting to ask ourselves what then happens to the spectrum 
with the double well potential case as an example as we discussed 
in the preceding subsection. 

It is clear that there should not be any continuum spectrum arising from 
a massless boson since there is no physical massless boson in two dimensions. 
Does this imply that the appearance of the massless boson $Hamiltonian$ is forbidden 
after the spontaneous symmetry breaking ? 

This is not a very easy question to answer since clearly the vacuum should prefer 
the lower energy state to the symmetric vacuum. What one can say with confidence is 
that the massless boson cannot become a physical particle in two dimensions. 
Since the boson field $\xi$ should be coupled to the other boson field $\rho$ 
in higher order interaction Hamiltonian, one may not conclude that the vacuum 
should be found at the symmetry preserving or broken state, before one obtains 
the spectrum of this boson field theory model by solving it exactly. 

As discussed in detail in section 2, the Goldstone theorem states 
that there should be a bosonic state which has the energy dispersion 
with $E=0$ for $\bm{p}=0$ when the symmetry is spontaneously broken. 
It has been believed that this state should be a massless boson. 
But this is, of course, a too strong statement. It only says that the dispersion 
of the state should be $E=0$ for $\bm{p}=0$, and could well be more complicated 
than the massless one, that is, $E=|\bm{p}| $ if the system is not an isolated one. 

In other words, the degrees of freedom of $\xi$ field may not become independent 
of the field $\rho$, and in this case, there is no reason to claim that there 
should appear a massless boson. It can be said that the state which satisfies 
the condition of $E=0$ for $\bm{p}=0$ does not necessarily correspond to a massless 
boson unless it is an independent field. If this state couples to other fields, 
then this complex field can survive free from the infra-red singularity of the massless 
boson propagator. In this sense, the condition of $E=0$ for $\bm{p}=0$ is not 
sufficient to forbid the existence of the symmetry broken vacuum.  

Therefore, it is still an open question what kind of spectrum 
should emerge from the boson field theory model with the double well potential 
in two dimensions. We believe that the spectrum in this boson field theory 
model should help us understand the symmetry breaking in two dimensions in depth.


\section{ No Goldstone Boson in Fermion Field Theory }

It has been believed that, when the symmetry of the Lagrangian density is 
spontaneously broken in the vacuum state, there should appear a massless boson 
in the fermion field theory models in the same way as the boson field theory.  
This belief started from the original work by Nambu and Jona-Lasinio (NJL) 
who studied the current current interaction model of the fermion field theory. 
By now, it is called the NJL model and it has the chiral symmetry. In their study, 
they showed that the vacuum prefers the chiral symmetry broken state and there should 
appear a massless boson associated with the chiral symmetry breaking.  However, 
they calculated the boson mass by summing up one loop Feynman diagrams based 
on the perturbative vacuum state. But it is obvious that the field theory 
calculation should be based on the physical vacuum state, and the physical 
vacuum in this NJL model is of course the new vacuum that breaks the chiral 
symmetry. Therefore, any physical observables should be evaluated starting 
from the symmetry broken vacuum. Otherwise one obtains a boson mass which 
is unphysical, and this unphysical massless boson is exactly what Nambu and 
Jona-Lasinio obtained. The reality is that there is no massless boson, and 
in addition there should be no boson in the NJL model even though 
the latter claim is not proved yet. 

In the NJL model, the massless fermion acquires an induced mass when we employ 
the Bogoliubov transformation method. Thus, the NJL model becomes a massive 
fermion field theory after the spontaneous symmetry breaking. This effective 
fermion mass was supposed to be the nucleon mass in the original paper 
of Nambu and Jona-Lasinio. However, we believe that this mechanism of 
the effective fermion mass is spurious, and it is only due to the approximation 
of the Bogoliubov transformation method. The spontaneous symmetry breaking 
phenomena can arise from the change of the vacuum state of the field theory 
model, and this should not change any property of the elementary fermion 
field itself. This is just in contrast to the boson field theory where 
the property of the boson field is basically determined by the boson mass, 
and bosons can be easily created or destroyed. But one cannot create 
any fermions since they are fundamental particles and there is no way 
to induce the mass scale for the massless fermion from the renormalization 
procedure. This point can be seen quite nicely in the Bethe ansatz solution 
in the massless Thirring model, and we will discuss it later. 

\subsection{ Intuitive Discussion}

Here, we present an intuitive discussion of the chiral symmetry 
breaking in the NJL models and show that there should not appear 
any massless boson at all \cite{q91}. The treatment here is far from 
rigorous, but we believe 
that the essential physics of the spontaneous symmetry breaking phenomena and 
bosons associated with the symmetry breaking in fermion field theory models 
should be clarified since there is still a  misunderstanding in this problem.  
The treatment is somewhat similar to the Bogoliubov transformation method 
which is originally employed by Nambu and Jona-Lasinio when they calculated the vacuum energy 
after the symmetry breaking in their model. Here, the determination of the vacuum energy is done 
by an educated guess although the result is quite similar to the one which is obtained 
by the Bogoliubov transformation method. 

The Hamiltonian density of the NJL model is written as 
$$ {\cal H}= \psi^{\dagger} \bm{p}\cdot \bm{\alpha}\psi -{1\over 2}G \biggl[ (\bar{\psi}\psi )^2 
+(\bar{\psi}i\gamma_5\psi )^2  \biggr] . \eqno{(4.1)}  $$
Here, we take the chiral representation, and denote the $\psi$ as 
$$ \psi(\bm{n},s)=\frac{1}{\sqrt{2}}
\left( \begin{array}{c}
 \psi_1  \chi^{(s)} \\
      \psi_2 \chi^{(s)} \end{array}
\right) \eqno{(4.2)} $$
where  $\chi^{(s)}$ denotes the spin part. 
In this case, the Hamiltonian density for the NJL model 
can be written after the summation of $s$ is taken 
$$ {\cal H}= \psi_1^{\dagger} (\bm{p}\cdot \hat{\bm{n}}) \psi_1 
-\psi_2^{\dagger} (\bm{p}\cdot \hat{\bm{n}}) \psi_2
 + 2G ( \psi_1^{\dagger} \psi_1 \psi_2^{\dagger} \psi_2 ) .  \eqno{(4.3)}  $$
In the same way as the boson case, we can define the potential $U(\psi_1,\psi_2)$ as 
$$ U(\psi_1,\psi_2) = 2G|\psi_1|^2|\psi_2|^2 . \eqno{(4.4)} $$
It is clear from this equation that the potential of the fermion 
field theory models does not have any nontrivial minimum, apart from the trivial one 
$ \psi_1^{\dagger} \psi_1=0$, $\psi_2^{\dagger} \psi_2=0$. 
This is in contrast to the boson case where there is a 
nontrivial minimum in the potential. 
Therefore, there is no degeneracy of the true vacuum state since the minimum 
of the potential here is only a trivial one. 

Where can one find the new vacuum that breaks the chiral 
symmetry ? The answer is simple. One has to consider the kinetic 
energy term. In the fermion system, the kinetic energy is negative 
for the vacuum state. 

In what follows, we present a simple and intuitive argument of obtaining a new 
vacuum state including the kinetic energy term. This treatment is schematic, 
but one can learn the essence of the physics of the chiral symmetry 
breaking in fermion field theory. The treatment by employing the Bogoliubov 
transformation method will be given in the next section. 
Now, we can take an average value  
of the kinetic energy ($-\Lambda_0$) for the negative energy state, 
and thus we write eq.(4.3) as
$$ {\cal H} \approx -\Lambda_0 \left(|\psi_1|^2 +|\psi_2|^2 \right)
 + 2G  |\psi_1|^2 | \psi_2 |^2  .  \eqno{(4.5)}  $$
This can be rewritten as 
$$ {\cal H}=  2G \left( |\psi_1|^2-{\Lambda_0\over{2G}} \right) 
\left( |\psi_2|^2-{\Lambda_0\over{2G}} \right)
-{\Lambda_0^2\over{2G}} .   \eqno{(4.6)}  $$
Therefore, it is easy to find the $|\psi_1|^2$ and 
$|\psi_2|^2 $ for the new vacuum state, that is, 
$$  |\psi_1|^2=  {\Lambda_0\over{2G}},  \qquad
  |\psi_2|^2=  {\Lambda_0\over{2G}}  . \eqno{(4.7)} $$
This result is somewhat similar to the mean field approximation and 
indeed the mean field approximation gives rise to the chiral symmetry breaking. 

In this case, the vacuum energy $E_{vac}$ 
and the condensate $C$ become 
$$ E_{vac}= -{\Lambda_0^2V\over{2G}} \eqno{(4.8a)}   $$
$$ C={\Lambda_0\over G}  \eqno{(4.8b)} $$
where $V$ denotes the volume of the system. 
Now, we write the $\psi_1$ and $\psi_2$ as
$$  \psi_i=\psi_0 + \tilde{\psi_i},  \quad (i=1,2) \eqno{(4.9)} $$
where $\psi_0$ denotes the fermion field which is assumed to satisfy 
the following relations
$$ \psi^*_0 \psi_0 ={\Lambda_0\over{2G}}   \eqno{(4.10a)} $$
$$ \psi_0\psi_0 =0   .  \eqno{(4.10b)} $$ 
In this case, the Hamiltonian density becomes
\begin{align*}
 {\cal H} = -{\Lambda_0^2\over{2G}} + 
\Lambda_0\left( \tilde{\psi}_1^\dagger\tilde{\psi_2} +h.c. \right)
&+ \tilde{ \psi}_1^{\dagger} (\bm{p} \cdot \hat{\bm{n}}) \tilde{ \psi}_1 
 -\tilde{\psi}_2^{\dagger} (\bm{p}\cdot \hat{\bm{n}})  \tilde{\psi}_2 \\
&+ 2G  |\tilde{\psi}_1|^2 |\tilde{ \psi}_2 |^2+ 
O(\tilde{ \psi}_1, \tilde{\psi}_2^{\dagger}) .   \tag{4.11} 
\end{align*}
Now, it is clear that the second term is the mass term. Therefore, one notices 
that after the chiral symmetry breaking, the fermion acquires the finite mass, 
and the induced mass $M$ becomes $M=\Lambda_0$. Therefore, at this point, 
the symmetry breaking problem is completed. The rest of the field theory becomes 
just the massive fermion field theory. For example, the Thirring model becomes 
the massive Thirring model where one knows well that there exists one massive boson, 
and the mass spectrum is obtained as the function of 
the coupling constant \cite{q13,q12,q14}. 

This means that one cannot find a massless 
boson in the Hamiltonian of the fermion system. It is also quite important 
to note that the new Hamiltonian is still described by the same number 
of the fermion degrees of freedom as the original one. 
This is in contrast to the boson case 
where one of the complex field freedom becomes the massless boson $\xi$. 

Therefore, if one wants to find any boson in the NJL model, 
then one has to solve the dynamics since the kinematics cannot produce any Goldstone boson 
in fermion field theory. However, it is difficult to find 
a massless boson as a bound state of fermion and antifermion system, 
regardless the strength of the coupling constant in the system 
of the finite fermion mass.  In any fermion field theory models, the bound 
state energy should depend on the strength of the interaction, and if 
there exists a massless boson in the massive fermion field theory model, 
this must be the strong coupling limit of the interaction strength. However, 
Nambu claimed the existence of a massless boson, regardless the strength of 
the coupling constant, and one can easily see that this claim is physically 
out of question. 

IndeedCthere is no massless boson in the NJL as well as 
in the massless Thirring models if one solves the dynamics properly 
as will be seen below in the next section. 

Here,  it is interesting to note that the vacuum energy 
and the condensate with this value of the $\Lambda_0$ [eq.(4.8$a$) and eq.(4.8$b$)] 
become 
$$ E_{vac}= -{M^2V\over{2G}}  \eqno{(4.12a)} $$  
$$ C={M\over G} \eqno{(4.12b)} $$ 
which are quite close to the Bogoliubov transformation calculations. 
Also, the chiral charge $Q_5$ of the vacuum can be evaluated and is found to be 
$$ Q_5 =0  \eqno{(4.13)} $$
which is also consistent with the one calculated by the Bogoliubov 
transformation method.

We summarize the intuitive discussions here for the fermion field 
theory. Even though there is no nontrivial minimum in the potential, 
one finds a new vacuum if one considers the kinetic energy terms of the negative energy 
particles in the vacuum state. 
The chiral symmetry is broken in the new vacuum state 
of the NJL and the Thirring  models. 
Namely, the momentum distributions of the negative energy particles 
in the vacuum states are rearranged such that the new vacuum energy  
becomes lower than the perturbative vacuum state. 
In this process, the left and right moving fermions change the momentum distributions 
in the vacuum state such that the chirality is broken since the broken state has 
simply a lower energy than the symmetry preserving vacuum state. 
After the symmetry breaking, the massless fermion acquires the effective mass 
though it is an approximate scheme. But there is no massless boson 
since the degree of freedom for the massless boson does not exist. 
The mass of the boson predicted in the field theory of the finite fermion mass 
has therefore nothing to do with the symmetry breaking business 
since it is just the same as asking for the excitation spectrum of 
the field $\rho$ in eq.(3.4).

\subsection{Bogoliubov Transformation} 

Now, we carry out the calculation which is based on the Bogoliubov 
transformation, and show that the chiral symmetry is indeed broken.  
However, we also show that there appears no massless boson in this 
regularized NJL model.

Here, we first quantize the fermion field in a box $L^3$ 
$$ \psi(\bm{r}) = \frac{1}{\sqrt{L^3}}\sum_{\bm{n},s}
\left[a_{\bm{n},s} u(\bm{n},s)e^{i{2\pi\over L} \bm{n} \cdot \bm{r}}+
b^{\dagger}_{\bm{n},s} v(\bm{n},s) e^{-i{2\pi\over L} \bm{n}\cdot \bm{r}}\right] 
\eqno{ (4.14)} $$
where $s$ denotes the spin index, and $s= \pm 1$. Also, the spinors are defined as 
$$ u(\bm{n},s)=\frac{1}{\sqrt{2}}
\left( \begin{array}{c}
 \bm{\sigma} \cdot \hat{\bm{n}}\chi^{(s)} \\
       \chi^{(s)} \end{array}
\right), \quad $$
$$ v(\bm{n},s)=\frac{1}{\sqrt{2}}
\left(  \begin{array}{c}
\chi^{(s)} \\
 \bm{\sigma} \cdot \hat{\bm{n}}\chi^{(s)} \end{array}
\right) .  $$
Now, we define new fermion operators by the Bogoliubov transformation, 
$$ c_{\bm{n},s} = e^{- {\cal A} } a_{\bm{n},s} e^{{\cal A}} = 
\cos \theta_{\bm{n}}  a_{\bm{n},s}
+ s \sin \theta_{\bm{n}} b^{\dagger}_{-\bm{n},s} \eqno{(4.15a)} $$
$$ d^{\dagger}_{-\bm{n},s} = e^{- {\cal A} } b^{\dagger}_{-\bm{n},s} e^{{\cal A}} 
= \cos \theta_{\bm{n}} b^{\dagger}_{-\bm{n},s} - 
s \sin \theta_{\bm{n}} a_{\bm{n},s} \eqno{(4.15b)} $$
where the generator of the Bogoliubov transformation is given by
$$ {\cal A} = \sum_{\bm{n},s} s \theta_{\bm{n}}
 \left( a^{\dagger}_{\bm{n},s} b^{\dagger}_{-\bm{n},s}  -b_{-\bm{n},s} a_{\bm{n},s} \right)  .
 \eqno{(4.16)} $$
$ \theta_{\bm{n}}$ denotes the Bogoliubov angle 
which can be determined by the condition that the vacuum energy is minimized. 
In this case, the new vacuum state is obtained as 
$$ \mid \Omega \rangle = e^{ - {\cal A} }  |0\rangle  .  \eqno{(4.17)}  $$
In what follows, we treat the NJL Hamiltonian with the Bogoliubov transformed 
vacuum state. In order to clearly see some important difference between 
the massive fermion and massless fermion cases, we treat the two cases 
separately. 

\subsection{Massive Fermion Case}

The Lagrangian density for the NJL model with the massive fermion can be written as 
$$ {\cal L}= i \bar \psi  \gamma_{\mu} \partial^{\mu}  \psi  -m_0\bar{\psi}\psi 
+G \left[ (\bar{\psi}\psi )^2 
+(\bar{\psi}i\gamma_5\psi )^2  \right]  . \eqno{(4.18)}  $$
Now, we can obtain the new Hamiltonian under the Bogoliubov transformation, 
\begin{multline*}
H = \sum_{\bm{n},s} \left\{|\bm{p}_{\bm{n}}| \cos 2\theta_{\bm{n}} 
+\left(m_{0}+\frac{2G}{L^3}\mathcal{B}\right) \sin 2\theta_{\bm{n}} \right\}
\left( c^{\dagger}_{\bm{n},s} c_{\bm{n},s} +d^{\dagger}_{-\bm{n},s} d_{-\bm{n},s}
\right) \\
 + \sum_{\bm{n},s}
\left\{-|\bm{p}_{\bm{n}}|s \sin 2\theta_{\bm{n}}+
\left(m_{0}+\frac{2G}{L^3}\mathcal{B}\right)s \cos 2\theta_{\bm{n}} \right\}
\left(c^{\dagger}_{\bm{n},s} d^{\dagger}_{-\bm{n},s} + d_{-\bm{n},s} c_{\bm{n},s} \right) 
 + {H'}_{int}  \tag{4.19}
\end{multline*}
where $ {H'}_{int}$ is the interaction term. Since the ${H'}_{int}$ 
is quite complicated, and 
besides its explicit expression is not needed in this context, we will not 
write it here.   
${\mathcal{B}} $ is defined as 
$$ {\mathcal{B}}=\sum_{{\bf n},s}\sin 2\theta_{\bm{n}} . $$ 
Now, we can define the renormalized fermion mass $m$
$$ m = m_{0}+\frac{2G}{L^3}\mathcal{B}. \eqno{ (4.20)} $$
The Bogoliubov angle $\theta_{\bm{n}}$ can be determined by imposing the condition 
that the $cd$ term in eq.(4.19) must vanish. Therefore, we obtain
$$ \cot 2\theta_{\bm{n}}  = {|\bm{p}_{\bm{n}}|\over m} . \eqno{ (4.21)} $$
This Bogoliubov angle $\theta_{\bm{n}}$ does not change when the mass 
varies from $m_0$ to $m$. 
In this case, the vacuum is just the same as the trivial vacuum of the massive case, 
except that the fermion mass is replaced by the renormalized 
mass $m$. The rest of the theory 
becomes identical to the massive NJL model with the same  
interaction Hamiltonian ${H'}_{int}$. Therefore, there is no symmetry breaking, 
and this vacuum has no condensate. 

\subsection{Massless Fermion Case}

Here, we present the same procedure for the massless fermion 
case in order to understand why the fermion has to become massive. 

We start from the Lagrangian density with no mass term in eq.(4.18). 
Under the Bogoliubov transformation, we obtain the new Hamiltonian
\begin{align*}
H &= \sum_{\bm{n},s}\left\{|\bm{p}_{\bm{n}}| \cos 2\theta_{\bm{n}} 
+\frac{2G}{L^3}\mathcal{B} \sin 2\theta_{\bm{n}}\right\}
\left( c^{\dagger}_{\bm{n},s} c_{\bm{n},s} + 
d^{\dagger}_{-\bm{n},s} d_{-\bm{n},s} \right) \\
 & + \sum_{{\bf n},s}
\left\{-|\bm{p}_{\bm{n}}| s \sin 2\theta_{\bm{n}}+
\frac{2G}{L^3}{\mathcal{B}} s \cos 2\theta_{\bm{n}} \right\}
\left(c^{\dagger}_{\bm{n},s} d^{\dagger}_{-\bm{n},s} + d_{-\bm{n},s} c_{\bm{n},s} \right) 
+ {H'}_{int}  \tag{4.22}
\end{align*}
where ${H'}_{int}$ is just the same as the one given in eq.(4.19). 
From this Hamiltonian, we get to know that the mass term is 
generated in the same way as the massive case. But we cannot 
make any renormalization since there is no mass term. Further, 
the new term is the only mass scale in this Hamiltonian since 
the coupling constant cannot serve as the mass scale. 
In fact, it is even worse than the dimensionless coupling constant case, 
since the coupling constant in the NJL model is proportional to the inverse 
square of the mass dimension. 
Thus, we define the new fermion mass $M_N$ by
$$ M_N = \frac{2G}{L^3}\mathcal{B}. \eqno{ (4.23)} $$
The Bogoliubov angle $\theta_{\bm{n}}$ can be determined from the following 
equation
$$ \cot 2\theta_{\bm{n}}  = {|\bm{p}_{\bm{n}}|\over M_N} . \eqno{ (4.24)} $$
In this case, the vacuum changes drastically since the original 
vacuum is trivial. 
 
Further, the constraints of eqs.(4.23) and (4.24) give rise to the equation 
that determines the relation between the induced fermion mass $M_N$ 
and the cutoff momentum $\Lambda$   
$$ M_N={4G\over{(3\pi)^3}}\int^{\Lambda}d^3p {M_N\over{\sqrt{M_N^2+p^2}}} .
\eqno{(4.25)} $$
This equation has a nontrivial solution for $M_N$, and the vacuum 
energy becomes lower than the trivial vacuum ($M_N =0$). 
Therefore, $M_N$ can be expressed in terms of $\Lambda$ as
$$ M_N=\gamma \Lambda $$
where $\gamma$ is a simple numerical constant. 

It should be noted that the treatment up to now is exactly the same as 
the one given by Nambu and Jona-Lasinio \cite{q7}. Further, we stress 
that the induced fermion mass $M_N$ can never be set to zero, 
and it is always finite. 

\subsection{Boson Mass in NJL Model}

The boson state $|B\rangle$  can be expressed as
$$ |B\rangle = \sum_{\bm{n},s} f_{\bm{n}}
c^{\dagger}_{\bm{n},s} d^{\dagger}_{-\bm{n},s} |\Omega\rangle, 
 \eqno{(4.26)} $$
where $f_{\bm{n}}$ is a wave function in momentum space, and $|\Omega\rangle$ 
denotes the Bogoliubov vacuum state.
The equation for the boson mass $ {\cal M}$ for the NJL model  is written 
in terms of the Fock space expansion at the large $L$ limit 
$$ {\cal M}f(p) = 2E_{p}f(p)
-\frac{2G}{(3\pi)^3}\int^{\Lambda} d^3q  f(q)\left(1+
\frac{M^2}{E_{p}E_{q}}+\frac{ \bm{p} \cdot \bm{q}}{E_{p}E_{q}}\right)  
 \eqno{(4.27)} $$
where  $M$ should be taken to be $M=m$ for the massive case, and 
$M=M_N$ for the massless case.  It is important to note that the fermion 
mass $M$ after the Bogoliubov transformation, therefore, cannot become zero. 

Here, again, we note that the RPA 
calculation gives the similar boson spectrum  
to the Fock space expansion. But we do not know whether the RPA calculation 
is better than the Fock space expansion or not, since the derivation 
of the RPA equation in field theory is not based on the fundamental 
principle. In principle, the RPA calculation may take into account the effect 
of the deformation of the vacuum in the presence of the particle and antiparticle. 
However, this is extremely difficult to do it properly, and indeed the RPA 
eigenvalue equation is not Hermite, and thus it is not clear whether 
the effect is taken into account in a better way or worse. 
The examination and the validity of the RPA equation 
will be given else where. 

The solution of eq.(4.25) can be easily obtained, and the boson mass spectrum 
for the NJL model is shown 
in Fig. 1. Note that the boson mass is measured in units of the cutoff momentum $\Lambda$. 
As can be seen from the figure, there is a massive boson for some regions 
of the values of the coupling constant. Here, as we will see later, the NJL 
and the Thirring models are quite similar to each other. 
This is mainly because the current-current interaction is essentially a delta 
function potential in coordinate space. 
Indeed, as is well known, the delta function potential in one dimension can always bind 
the fermion and anti-fermion while the delta function potential in three dimension 
cannot normally bind them. Due to the finite cut off momentum, 
the delta function potential in three dimensions 
can make a weak bound state, depending on the strength of the coupling constant. 
This result of the delta function potential in quantum mechanics is 
almost the same as  what is just shown in Fig. 1. 

%
%

\begin{figure}[htbp]
\centering
\includegraphics[width=0.7\textwidth]{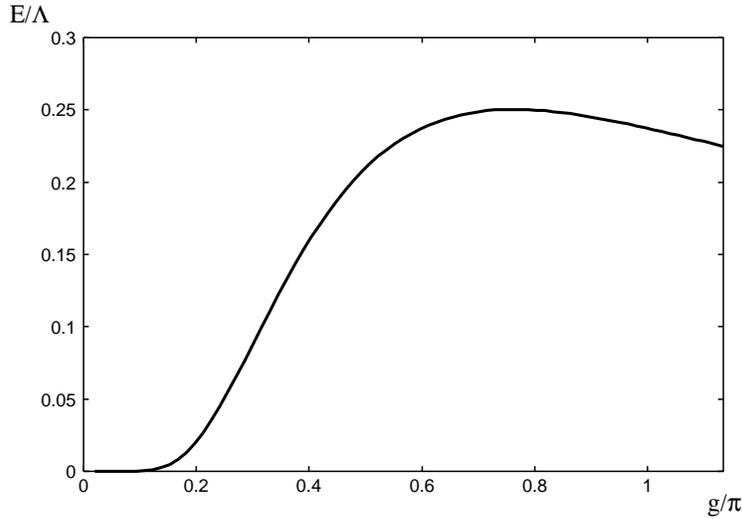}
\caption{The boson mass for the NJL model is plotted as the function of $G\Lambda^2$. 
It is measured by the cutoff $\Lambda$.}
\label{fig1}
\end{figure}

Further, we should note that Kleinert and Van den Bossche \cite{q7p} also found 
that the bosons after the symmetry breaking are all massive, which is just 
consistent with our claim. Their method and approach are quite different from 
the present calculation, but both of the calculations agree with each other 
that there is no massless boson in the NJL model. 

Here, we should add that there is no serious difficulty of proving 
the non-existence of the massless boson from the calculated spectrum. 
However, if it were to prove 
the existence of the massless boson, it would have been extremely difficult 
to do it. For the massless boson, there should be a continuum spectrum, and 
this continuum spectrum of the massless boson should be differentiated from 
the continuum spectrum arising from the many body nature of the system. 
This differentiation must have been an extremely difficult task without having 
some analytic expression of the spectrum. 
In fact, even if one finds a continuum spectrum which has, for example, 
the dispersion of $E=c_0 p^2$ as often discussed in solid state physics, 
one sees that the spectrum has nothing to do with the Goldstone boson.

\subsection{Boson Mass in Thirring Model}

The massless Thirring model can be treated just in the same way as the NJL model 
in terms of the Bogoliubov transformation method. We do not repeat the detailed 
calculations, but instead we present the summary of the calculated results. 

First, we can determine the Bogoliubov angles and from the consistency condition 
we can determine the induced mass.  
The induced mass $M$ can be expressed in terms of the cutoff $\Lambda$,
$$ M={\Lambda\over{\sinh \left({\pi\over g}\right)}} . \eqno{(4.28)} $$ 
Further, the vacuum energy $E_{vac}$ as measured from the trivial vacuum  is given 
$$ E_{vac}=-{L\over{2\pi}}{\Lambda^2\over{\sinh \left({\pi\over g}\right) }}
e^{-{\pi\over g}} . \eqno{(4.29)} $$
From this value of the vacuum energy, we get to know that the new vacuum energy 
is indeed lower than the trivial one. Therefore, the chiral symmetry is broken 
in the new vacuum state and, effectively, the fermion becomes massive. 

In the same manner as \cite{q9}, we carry out the calculations of the spectrum 
of the bosons in the Fock space expansion. 
The boson state $|B\rangle$  can be expressed as
$$ |B\rangle = \sum_{n}f_{n}c_{n}^{\dagger}d_{-n}^{\dagger}|\Omega\rangle,
 \eqno{(4.30)} $$
where $f_{n}$ is a wave function in momentum space, and $|\Omega\rangle$ 
denotes the Bogoliubov vacuum state. The energy eigenvalue 
of the Hamiltonian for the large $L$ limit can be written as 
$$ {\cal M}f(p) = 2E_{p}f(p)
-\frac{g}{2\pi}\int dq f(q)\left(
1+\frac{M^2}{E_{p}E_{q}}+\frac{pq}{E_{p}E_{q}}\right) 
 \eqno{(4.31)} $$
where $ {\cal M}$ denotes the boson mass. 
$E_{p}$ is given as 
$$ E_{p} = \sqrt{M^2+p^2} . \eqno{(4.32)}  $$
Eq.(4.31) can be solved exactly as shown in \cite{q9}. First, we define $A$ and $B$ by
$$ A = \int_{-\Lambda}^{\Lambda} dp f(p)  \eqno{(4.33a)}  $$ 
$$ B = \int_{-\Lambda}^{\Lambda} dp \frac{f(p)}{E_{p}}. \eqno{(4.33b)} $$
Using $A$ and $B$, we can solve Eq. (4.31) for $f(p)$ and obtain
$$ f(p) = \frac{g/2\pi}{2E_{p}-\mathcal{M}}\left(A+\frac{M^2}{E_{p}}B\right).
 \eqno{(4.34)} $$
Putting this $f(p)$ back into Eqs. (4.31), we obtain the matrix equations
$$ A = \frac{g}{2\pi}\int_{0}^{\Lambda} \frac{2dp}{2E_{p}-\mathcal{M}}
\left(A+\frac{M^2}{E_{p}}B\right) \eqno{(4.35a)}  $$
$$ B = \frac{g}{2\pi}\int_{0}^{\Lambda} \frac{2dp}{(2E_{p}-{\mathcal{M}})E_{p}}
\left(A+\frac{M^2}{E_{p}}B\right). \eqno{(4.35b)} $$
Since the model is already regularized, we can easily calculate the boson 
spectrum which is given in Fig. 2 as the function of the coupling 
constant $g/\pi$. 
As can be seen from Fig. 2, there is no massless boson 
in this spectrum even though the boson mass for the very small coupling constant $g$ 
is exponentially small.

%
%

\begin{figure}[htbp]
\centering
\includegraphics[width=0.7\textwidth]{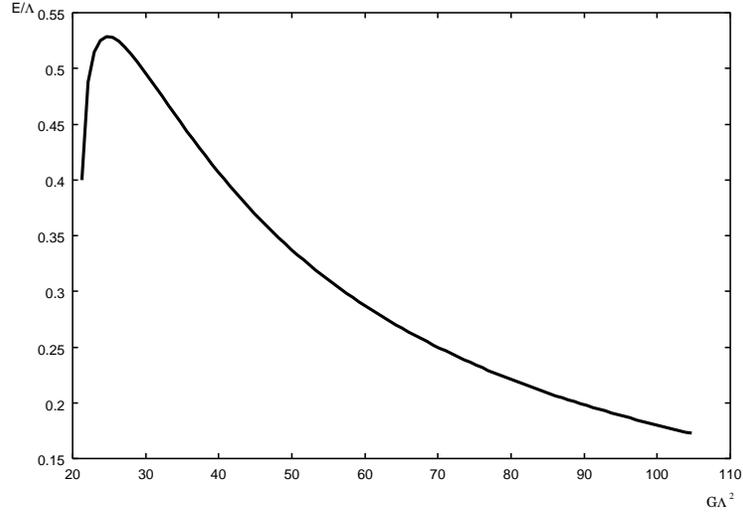}
\caption{The boson mass for the massless Thirring model is plotted 
as the function of $g/\pi$. It is measured by the cutoff $\Lambda$.}
\label{fig2}
\end{figure}


\section{ Bethe Ansatz Solutions in Thirring Model }

The structure of the vacuum and the spectrum of the current current 
interaction models of the NJL and the Thirring model have been discussed 
in terms of the Bogoliubov transformation method in section 4. It is clear 
that there should not be any fundamental differences between the NJL 
and the Thirring models  as far as the vacuum structure is concerned 
once one accepts the reliability of the Bogoliubov transformation method. 
For the perturbative point of view, however, there is a serious difference 
between them since the former is an unrenormalizable field theory model 
in four dimensions while the latter is a renormalizable field theory model 
in two dimensions. Therefore, there is no guarantee that the NJL model 
can give any reliable predictions when one evaluates physical observables 
in the perturbative calculations. In this respect, it is always very important 
to have some exact solutions of the field theory model. In two dimensions, 
the Thirring model can be indeed solved exactly by the Bethe ansatz method. 
From the exact solutions, one can learn a lot about the vacuum structure 
and the spectrum of the excitation in the Thirring model.  In particular, 
we can discuss the symmetry breaking of the vacuum state from this solution. 

In two dimensions, however, the symmetry in the field theory  
is considered to be $not$ broken spontaneously in the vacuum state. 
In fact, Coleman \cite{q3} presented the proof  
that the two dimensional field theory models cannot spontaneously break the symmetry 
even though the vacuum state may prefer the symmetry broken state. 
However, his proof of the nonexistence of the spontaneous symmetry breaking 
in two dimensions is essentially based on the Goldstone theorem. 
The Goldstone theorem \cite{q1,q2} states that the spontaneous symmetry breaking should 
accompany a massless boson when the vacuum prefers the broken symmetric 
state. However, the massless boson cannot exist in two dimensions since 
it cannot propagate due to the infra-red singularity of the propagator. 
Since this non-existence of the massless boson should hold  rigorously, 
it naturally means that the spontaneous symmetry breaking 
should not occur in two dimensions as long as the Goldstone theorem is right. 
Coleman's theorem looks reasonable, and indeed until recently 
it has been believed to hold true for fermion field theory models as well.  

However, as we see in the preceding sections, the Goldstone theorem 
does not hold in the fermion field theory models. 
Therefore, there is no massless boson in the fermion field theory  after 
the spontaneous symmetry breaking. This suggests that Coleman's theorem 
has lost its basis in the proof of the theorem. Indeed, 
as we saw in section 4, the chiral symmetry of the massless Thirring model is 
spontaneously broken by the Bogoliubov vacuum state \cite{q91,q9}. 
There, the energy of the new vacuum is lower than that of the free vacuum state, 
and it breaks the chiral symmetry, but there appears no massless boson. 

For this claim, however, people may insist that the Bogoliubov transformation 
does not have to be exact, and therefore there might be some excuse 
for the symmetry breaking phenomena that occurred in the Thirring model. 

In this section, we present a new discovery of the symmetry broken vacuum 
of the Bethe ansatz solution in the Thirring model, 
and show that the energy of the new vacuum state 
is indeed  lower than that of the symmetric vacuum state 
even though the symmetric vacuum was considered to be the lowest state 
in the Thirring model. The new vacuum state breaks the chiral symmetry, 
and becomes a massive fermion field theory model. 

Further, we evaluate the energy spectrum of the one particle-one hole states, 
and show that the excitation spectrum has indeed a finite gap. This gap energy 
turns out to be consistent with the effective fermion mass deduced  
from the momentum distribution of the negative energy particles 
in the new vacuum state. This confirms the consistency of the calculation 
of the Bethe ansatz solutions in the Thirring model. 

After carrying out the numerical calculations, we get to know that 
the energies of the vacuum  as well as the lowest one particle-one hole 
state can be expressed analytically. This is quite nice 
since we know clearly which of the vacuum state is the lowest. 
Also, in the thermodynamic limit, the lowest one particle-one hole 
state can be reduced to the effective fermion mass $M_N$ which is described 
in terms of the cutoff $\Lambda$. 

It turns out that there is no massive boson in the Bethe ansatz solutions, 
contrary to the prediction of the Bogoliubov transformation method \cite{q9,p5}. 
However, qualitative properties of the symmetry breaking phenomena 
between the Bethe ansatz calculations and the Bogoliubov method agree with 
each other.

Even though the  Bethe ansatz calculations confirm that there is 
no massless boson in the massless Thirring model, 
some people may claim that the massless 
Thirring model can be bosonized and is reduced to a massless boson 
Hamiltonian. Here, we show that the well-known procedure of bosonization 
of the massless Thirring model is incomplete because the zero mode 
of the boson field cannot be defined and quantized. In other words, 
the zero mode of the field $\Phi (0)$ identically vanishes 
in the massless Thirring model. This is in contrast to the Schwinger model 
in which one finds the zero mode of the field $\Phi (0)$ 
by the gauge field $A^1$. Also, it is interesting to note that the massive 
Thirring model has the zero mode through the mass term, and this clearly 
indicates that the massless limit of the massive Thirring model is indeed 
a singular point with respect to the dynamics of the field theory. 

Therefore, the massless Thirring model cannot be reduced to a free 
massless  boson even though it has a similar mathematical structure to 
the massless boson. The spectrum of the massless Thirring model has a finite gap,  
and this is consistent with the fact that there should not be any 
physical massless boson in two dimensions. Even though the defect 
of the bosonization of the massless Thirring model is only one 
point of the boson field, that is, zero mode, it is interesting 
and surprising that $nature$ knows it in advance.  

\subsection{Thirring Model and Bethe Ansatz  Solutions}
The massless Thirring model is a 1+1 dimensional field theory with 
current current interactions \cite{q7}. 
Its Hamiltonian can be written as 
$$H = \int dx \left\{-i\left(\psi_1^{\dagger}{\partial\over{\partial x}}\psi_1
-\psi_2^{\dagger}{\partial\over{\partial x}}\psi_2 \right)
+2g \psi_1^{\dagger}\psi_2^{\dagger}\psi_2\psi_1 \right\}.  \eqno{(5.1)}$$
The Hamiltonian eq.(5.1) can be diagonalized by the Bethe ansatz wave 
function for $N$ particles  \cite{q13,b5,b55}
$$\mid k_1, \cdots, k_N \rangle =\int dx_1 \cdots dx_{N_1}dy_1 \cdots dy_{N_2}
\prod_{i=1}^{N_1}  \exp (ik_ix_i) 
\prod_{j=1}^{N_2}  \exp \left(ik_{N_1+j}y_j\right) $$
$$ \times \prod_{i,j}\bigl[1+\lambda \theta (x_i-y_j) \bigr]
 \prod_{i=1}^{N_1}\psi_1^{\dagger}(x_i)
\prod_{j=1}^{N_2}\psi_2^{\dagger}(y_j)
\mid \! 0 \rangle, \eqno{(5.2)} $$
with $N_1+N_2 =N$. $\theta (x)$ denotes the step function. 
$k_i$ represents the momentum of the $i-$th particle. 
$ \lambda $ is determined to be \cite{q7}
$$ \lambda = -{g\over 2} S_{ij}  \eqno{(5.3)} $$
where $S_{ij}$ is defined as
$$ S_{ij}={k_iE_j-k_jE_i\over {k_ik_j-E_iE_j-\epsilon^2}}  \eqno{(5.4)} $$
where $\epsilon$ denotes the infra-red regulator which is eventually set to zero. 
We note that all of the momenta and any of the physical observables do not 
depend on the regulator $\epsilon$ when we solve the PBC equations as we discuss below.  

In this case, the eigenvalue equation becomes 
$$ H \mid k_1, \cdots, k_N \rangle  = \sum_{i=1}^N E_i  \mid
 k_1, \cdots, k_N \rangle  . \eqno{(5.5)} $$ 
From the periodic boundary condition (PBC), one obtains the following PBC equations,
$$ k_i = {2\pi n_i \over L} + {2 \over L}\sum_{j\not= i}^N 
\tan^{-1} \left({g\over 2} S_{ij} \right) \eqno{(5.6)} $$
where $n_i$'s are integer, and runs as $n_i=0, \pm 1,\pm 2, \cdots , N_0$ 
where
$$ N_0={1\over 2}(N-1). $$  

\subsection{Vacuum State}

First, we want to make a vacuum.  We write the PBC equations
for the vacuum which is filled with negative energy particles \cite{q13,b133}
$$ k_i = {2\pi n_i \over L} - {2 \over L}\sum_{\stackrel{\scriptstyle i\ne j}{k_i\ne k_j}}^N 
 \tan^{-1} \left({g\over 2} {k_i|k_j|-k_j|k_i|\over{k_ik_j-|k_i||k_j|-\epsilon^2 }} 
\right). \eqno{(5.7)} $$
Although, the expression of the phase shift function is somewhat different 
from that of ref. \cite{q42,q43}, it produces the same values of the momentum solution 
of the vacuum state. However, we believe that the expression of eq.(5.4) with 
the infra-red regulator must be better since it is transparent and clear. 

Here, we first fix the maximum 
momentum of the negative energy particles, 
and denote it by the cut off momentum $\Lambda$. Next, 
we take the specific value of $N$, and this leads to the determination of $L$ 
$$ L={2\pi N_0\over{\Lambda}} . \eqno{(5.8)} $$
If we solve eq.(5.7), then we can determine the vacuum state, 
and the vacuum energy $E_v$  can be written as
$$  E_v =- \sum_{i=1}^{N} |k_i| . \eqno{(5.9)} $$
It should be noted that physical observables are obtained by taking 
the thermodynamic limit where we let $L \rightarrow \infty$ and $N \rightarrow \infty$, 
keeping $\Lambda$ finite. If there is other scale like the mass, then one should 
take the  $\Lambda$ which is sufficiently large compared to the other scale. 
However, there is no other scale in the massless Thirring model or four dimensional QCD 
with massless fermions, and therefore all the physical observables are measured 
by the  $\Lambda$. Here, we can take all the necessary steps, if required,  since 
all the physical quantities are given analytically. In fact, as we see below, 
the excitation energy and the effective fermion mass are expressed in terms of 
the  $\Lambda$ in the thermodynamic limit. 

\subsection{Symmetric Vacuum State}

The solution of eq.(5.7) has been known and is written as \cite{q42,q43} 
$$ k_1=0 \eqno{(5.10a)} $$
for $n_1=0$,      
$$ k_i = {2\pi n_i \over L} + {2N_{0} \over L}
\tan^{-1} \left({g\over 2}  \right) \eqno{(5.10b)} $$
for $n_i=1,2,\cdots, N_0$,     
$$ k_i = {2\pi n_i \over L} - {2N_{0} \over L}
\tan^{-1} \left({g\over 2}  \right) \eqno{(5.10c)} $$
for $n_i=-1,-2,\cdots, -N_0$. This gives a symmetric vacuum state, 
and was considered to be the lowest state. 

The vacuum energy $E_v^{\rm sym}$ can be written as 
$$ E_v^{\rm sym}=-\Lambda \left\{ N_0+1+{2N_0\over \pi}
\tan^{-1} \left({g\over 2} \right) \right\} . \eqno{(5.11)} $$

\subsection{True Vacuum State}

It is surprising that eq.(5.7) has a completely different solution from the above 
analytical solutions. By the numerical calculation of eq.(5.7), 
we first find the new vacuum state. After that, we get to know that the solutions 
can be analytically written like the symmetric case,
$$ k_1={2N_0 \over L} \tan^{-1} \left({g\over 2}  \right) \eqno{(5.12a)} $$
for $n_1=0$,      
$$ k_i = {2\pi n_i \over L} + {2N_{0} \over L}
\tan^{-1} \left({g\over 2}  \right) \eqno{(5.12b)} $$
for $n_i=1,2,\cdots, N_0$,     
$$ k_i = {2\pi n_i \over L} - {2(N_{0}+1) \over L}
\tan^{-1} \left({g\over 2}  \right) \eqno{(5.12c)} $$
for $n_i=-1,-2,\cdots, -N_0$. The new vacuum  has no $k_i=0$ solution, 
and breaks the left-right symmetry.  Instead, all of the momenta of the 
negative energy particles become finite. 

The energy $E_v^{\rm true}$ of the true vacuum state can be written as 
$$ E_v^{\rm true}=-\Lambda \left\{ N_0+1+{2(N_0+1)\over \pi}
\tan^{-1} \left({g\over 2} \right) \right\} . \eqno{(5.13)} $$

From the distributions of the negative energy 
particles, one sees that this solution breaks the chiral symmetry. 
This situation can be easily seen from the analytical solutions 
since the absolute value of the momentum of the negative energy particles is 
higher than $\displaystyle{{\Lambda \over \pi} \tan^{-1} \left({g\over 2}  \right)} $. 
Therefore, we can define the effective fermion mass $M_N$ by 
$$ M_N = {\Lambda \over \pi} \tan^{-1} \left({g\over 2}  \right) . 
\eqno{(5.14)}  $$
In Table 1, we show the calculated results of the new vacuum as well as 
the symmetric vacuum energies as the function of  the particle 
number $N$. Here, we present the case with the coupling constant of $g= \pi $.  

\begin{table}
\caption{We show the calculated results of the vacuum energy of Bethe ansatz solutions 
at $ g=\pi $ with the particle number $N=401$ and $N=1601$.  
$\cal{E}_{\rm sym}$  and  $\cal{E}_{\rm true}$ denote the symmetric vacuum 
and the true vacuum energies, respectively. 
We also show the effective fermion mass ${\cal M}_N$ deduced from the vacuum momentum 
distributions. All the energies are measured in units of $\Lambda$, 
namely, ${\cal E} \equiv E/\Lambda $ and ${\cal M}_N \equiv M_N/\Lambda $.}
\begin{center}
\vspace{0.2cm}
\begin{tabular}{|c||c|c|c|}
\hline
\ $N$ & $\cal{E}_{\rm sym}$  & $\cal{E}_{\rm true}$  &  ${\cal M}_N$  \rule[-2mm]{0pt}{6mm}\\
\hline
\hline
401 & $-328.819$ & $-329.458$ & $0.320$  \\
\hline
1601 & $-1312.274$    & $-1312.913$ & 0.320  \\
\hline
\end{tabular} 
\end{center}
\end{table}


\subsection{$1p-1h$ State}

Next, we evaluate one particle-one hole $(1p-1h)$
 states. There, we take out one
negative energy particle ($i_0$-th particle)
 and put it into a positive energy state.
In this case, the PBC equations become
$$  k_i  = \frac{2\pi n_i}{L}-\frac{2}{L}\tan^{-1}
 \left( {g\over 2} {k_i|k_{i_0}|+k_{i_0}|k_i|\over{k_ik_{i_0}+|k_i||k_{i_0}|+\epsilon^2 }}  \right) $$
$$
 - \frac{2}{L}\sum_{\stackrel{\scriptstyle j\neq i,i_0}{k_j\ne k_i,k_{i_0}}}^N\tan^{-1}
\left({g\over 2} {k_i|k_j|-k_j|k_i|\over{k_ik_j-|k_i||k_j|-\epsilon^2 }} 
\right) \eqno{(5.15a)}$$
for $i\neq i_0$. $$  k_{i_0}  = \frac{2\pi n_{i_0}}{L}-
\frac{2}{L}\sum_{\stackrel{\scriptstyle j\neq i_0}{k_j\ne -k_{i_0}}}^N
 \tan^{-1}\left({g\over 2} {k_{i_0}|k_j|+k_j|k_{i_0}|
\over{k_{i_0}k_j+|k_{i_0}||k_j|+\epsilon^2 }} \right) \eqno{(5.15b)}$$
for $i= i_0$. In this case, the energy of the one particle-one
hole states $E^{1p1h}_{(i_0)}$ is given as,
$$  E^{1p1h}_{(i_0)}= |k_{i_0}| -
\sum_{\stackrel{\scriptstyle i=1}{i\not= i_0}}^{N}
|k_i|  .  \eqno{(5.16)}   $$
It turns out that the solutions of eqs.(5.15) can be found 
at the specific value of $n_{i_0}$ and then from this $n_{i_0}$ value  on, 
we find continuous spectrum of the $1p-1h$ states. 

Here, we show the analytical solution of eqs.(5.15) for the lowest $1p-1h$ state. 
$$ k_{i_0}={2\pi n_{i_0}\over L}-{2N_0\over L}\tan^{-1}\left({g\over 2}\right) \eqno{(5.17a)} $$  
for $n_{i_0}$,      
$$ k_i = {2\pi n_i \over L} + {2(N_{0}+1) \over L}
\tan^{-1} \left({g\over 2}  \right)\eqno{(5.17b)} $$
for $n_i=0,1,2,\cdots, N_0$    
$$ k_i = {2\pi n_i \over L} - {2N_{0} \over L}
\tan^{-1} \left({g\over 2}  \right)\eqno{(5.17c)} $$
for $n_i=-1,-2,\cdots,-N_0$. $n_{i_0}$ is given by 
$$ n_{i_0}=\left[ {N_0\over{\pi}}\tan^{-1} \left({g\over 2}\right)  \right] , 
\eqno{(5.18)} $$ 
where $[X]$ denotes the smallest integer value which is larger than $X$. 
In this case, we can express the lowest $1p-1h$ state energy analytically
$$ E_0^{1p-1h}=-\Lambda
\left\{(N_0+1)-{{2n_{i_0}}\over N_0}+{2(N_0+1)\over \pi}\tan^{-1}\left({g\over\pi}\right)
\right\}. \eqno{(5.19)} $$
Therefore, the lowest excitation energy $\Delta E_0^{1p-1h}$ 
with respect to the true vacuum state becomes 
$$ \Delta  E_0^{1p-1h} \equiv E_0^{1p-1h}-E_v^{\rm true}=
{2\Lambda\over N_0}  n_{i_0} . \eqno{(5.20)} $$
If we take the thermodynamic limit, that is, $ N\rightarrow \infty $ and  
$ L\rightarrow \infty $, then eq.(5.18) can be reduced to 
$$ \Delta  E_0^{1p-1h} ={2\Lambda \over \pi} \tan^{-1} 
\left({g\over 2}  \right)  = 2M_N  . \eqno{(5.21)} $$
In Table 2, we show the lowest five states of the $1p-1h$ energy by the numerical 
calculation. From this, we can determine the gap energy.

\begin{table}[h]
\caption{
We show several lowest states of the calculated results of  the 1p-1h states energy 
$\cal{E}$ of eqs.(4.12) at $ g=\pi  $ with $N=1601$.
The gap energy $\Delta {\cal E} \equiv {\cal E}^{(1p1h)}-{\cal E}_v$ is also shown. 
All the energies are measured in units of $\Lambda$. }

\begin{center}
\begin{tabular}{|c||c|c|}
\hline
\   & $\cal{E}$  &  $\Delta \cal{E} $  \rule[-2mm]{0pt}{6mm}\\
\hline
\hline
 vacuum  & $-1312.913$  & \   \\
\hline
${1p-1h}$ (1) & $-1312.273$ & $0.640$  \\
\hline
${1p-1h}$ (2)  & $-1312.272$ & $0.641$  \\
\hline
${1p-1h}$ (3)  & $-1312.271$ & $0.642$  \\
\hline
${1p-1h}$ (4)   & $-1312.269$ & $0.644$  \\
\hline
${1p-1h}$ (5)  & $-1312.268$ & $0.645$  \\
\hline
\end{tabular} 
\end{center}
\end{table}


From this gap energy, we can obtain the effective fermion mass which is one half 
of the lowest gap energy. This can be easily given as 
$$ M_N = 0.320 \  \Lambda . \eqno{(5.22)}  $$
This is consistent with the effective fermion mass 
deduced from the negative energy distribution of the vacuum. 
This confirms the consistency of the present calculations. 

\subsection{Boson State}

In this calculation, we do not find any boson state, contrary to the prediction 
of the Bogoliubov transformation method. Since the present calculation is exact, 
we believe that the Bogoliubov calculation overestimates the attraction 
between the particle hole states. The main difference between the Bethe solutions and 
the Bogoliubov vacuum arises from the dispersion relation 
of the negative energy particles. From the Bethe ansatz solutions, 
it is clear that one cannot make a simple free particle dispersion 
with the fermion mass term while the Bogoliubov method 
assumes the free fermion dispersion relation for the negative energy particles. 
This should generate slightly stronger attraction for the Bogoliubov vacuum state 
than for the Bethe ansatz solution. 

However, as far as the symmetry breaking mechanism is concerned, the Bogoliubov 
transformation gives a sufficiently reliable description of the dynamics 
in the spontaneous symmetry breaking phenomena. 

\subsection{Bosonization of Massless Thirring Model}

Here, we briefly review the bosonization procedure in two dimensional 
field theory models. In particular, we discuss the massless and massive Thirring models 
and show that the massless Thirring model cannot be bosonized properly due to the lack 
of the zero mode of the boson field. 

It has been believed that the massless 
Thirring model can be bosonized \cite{p33} in the same way as above, and its Hamiltonian is written 
$$ H= {1\over 2}\sum_{p \not= 0} \left\{ \left(1-{g\over{2\pi}} \right)  {\Pi}^{\dagger} (p) \Pi (p)
+\left( 1+{g\over{2\pi}} \right) p^2 {\Phi}^\dagger (p)\Phi (p)  \right\} . 
\eqno{(5,23)} $$
This looks plausible, but one knows at the same time 
that the $p=0$ part is not included. In fact, 
there is a serious problem in the definition 
of the boson field $\Phi (0)$ and $\Pi (0)$ at the zero momentum $p=0$. 
From eqs.(4.1), it is clear that one cannot define the zero mode 
of the boson field. In the Schwinger model, 
one finds the $\Phi (0)$ due to the anomaly equation. However, the Thirring 
model has no anomaly, and therefore the $\Phi (0)$ identically vanishes. 
That is, 
$$ \Phi (0) =0 .  \eqno{(5.24)} $$
There is no way to find the corresponding zero mode of the boson field 
in the massless Thirring model since the axial vector current is always conserved. 

Therefore, the Hamiltonian of the massless 
Thirring model eq.(5.5) does not correspond to the massless boson. 
It is interesting to notice 
that the problem is closely related to the zero mode which exhibits 
the infra-red property of the Hamiltonian. This is just consistent 
with the non-existence of the massless boson due to the infra-red 
singularity of the propagator in  two dimensions \cite{q4}. 
Further, as discussed in the previous section, the Bethe ansatz solutions 
confirm the finite gap of the massless Thirring spectrum, and 
this rules out a possibility of any excuse of the massless boson 
in the massless Thirring model.

\subsection{Physics of Zero Mode}

What is the physics behind the Hamiltonian without the zero mode ? 
Here, we discuss the effect of the zero mode and the eigenvalues of the Hamiltonian 
in a simplified way. The Hamiltonian eq.(5.23) can be rewritten as 
$$ H= H_B
-{1\over 2} \left( 1-{g\over{2\pi}} \right)  {\Pi}^{\dagger} (0) \Pi (0) \eqno{(5.25)} $$
where the $\Pi(0)$ field is introduced by hand, and the existence of 
the  $\Pi(0)$ and $\Phi(0)$  fields is assumed. 
Here, $H_B$ denotes the free boson Hamiltonian and is written as 
$$ H_B= {1\over 2}\sum_{p } \left\{ \left(1-{g\over{2\pi}} \right)  {\Pi}^{\dagger} (p) \Pi (p)
+\left(1+{g\over{2\pi}} \right) p^2 {\Phi}^\dagger (p)\Phi (p)  \right\} . \eqno{(5.26)} $$
Now, we assume the following eigenstates for $H_B$ and 
$ {\Pi}^{\dagger} (0) \Pi (0)$ by
$$ H_B|p\rangle =E_p|p\rangle \eqno{(5.27a)} $$
$$ {\Pi}^{\dagger} (0) \Pi (0) |\Lambda \rangle  =\Lambda |\Lambda \rangle \eqno{(5.27b)} $$
where $E_p={2\pi\over L} p  \ \ {\rm with} \ \ p=0,1,2, \cdots  $, and  
$\Lambda$ is related to the box length $L$ by $\Lambda ={c_0\over{L}}$ with $c_0$ constant. 

Eq. (5.27a) is just the normal eigenvalue equation for the massless boson and 
its spectrum. On the other hand, eq.(5.27b) is somewhat artificial 
since the state $ |\Lambda \rangle $ is introduced by hand. The zero mode state 
of the Hamiltonian $H_B$ should couple with the state $ |\Lambda \rangle $, and 
therefore new states can be made by the superposition of the two states
$$  |v \rangle = c_1  |\Lambda \rangle +c_2 
|0 \rangle  \eqno{(5.28)} $$
where $c_1$ and $c_2$ are constants. 
Further, we assume for simplicity that the overlapping integral between the $|0\rangle$ 
and the $|\Lambda \rangle$ states is small and is given by $\epsilon$
 $$ \langle 0|\Lambda \rangle =\epsilon . \eqno{(5.29)} $$
In this case, the energy eigenvalues $\langle v  |H |v \rangle$ of eq.(5.25) 
become at the order of $O(\epsilon)$
$$ E_{\Lambda} = \langle \Lambda  |H_B |\Lambda \rangle 
-{1\over 2} \left( 1-{g\over{2\pi}} \right) \Lambda \eqno{(5.30a)} $$
$$ E_0 = -{1\over 2} \left( 1-{g\over{2\pi}} \right) 
\langle 0  | {\Pi}^{\dagger} (0) \Pi (0)|0 \rangle . \eqno{(5.30b)} $$
If we assume that the magnitude of the $\langle \Lambda  |H_B |\Lambda \rangle $ and 
$\langle 0  | {\Pi}^{\dagger} (0) \Pi (0)|0 \rangle $ should be appreciably smaller than the $\Lambda$, 
$$ \langle \Lambda  |H_B |\Lambda \rangle \ll  \Lambda \eqno{(5.31a)}  $$ 
$$ \langle 0  | {\Pi}^{\dagger} (0) \Pi (0)|0 \rangle  \ll  \Lambda \eqno{(5,.31b)}  $$ 
then the spectrum of the Hamiltonian eq.(5.25) has a finite gap, and 
the continuum states of the massless excitations start right above the gap. This is 
just the same as the spectrum obtained from the Bethe ansatz solutions discussed 
in the previous section. 

\subsection{Bosonization of Massive Thirring Model}

It is well known that the massive Thirring model is equivalent 
to the sine-Gordon field theory \cite{col}. 
The proof of the equivalence is based on the observation that the arbitrary number 
of the correlation functions between the two models agree with each other 
if some constants and the fields of the two models are properly identified between them.  
This indicates that the massive 
Thirring model must be well bosonized. 

This is now quite clear since the axial vector current conservation 
is violated by the mass term,
$$ \partial_\mu j^{\mu}_5 = 2im \bar \psi \gamma_5  \psi 
\eqno{(5.32)} $$
where $j_{\mu}^5$ is defined as 
$$  j_{\mu}^5 = \bar \psi \gamma_5 \gamma_{\mu} \psi . \eqno{ (5.33)}   $$
It should be noted that the $j_{0}^5$ is equal to $j_{1}$ in two dimensions. 

Therefore, one can always define the ${\dot Q}_5$ by
$$ {\dot Q}_5 =2im \int \bar \psi \gamma_5  \psi dx . \eqno{(5.34)} $$
Therefore, one obtains the field $\Phi (0)$ of the boson 
$$ \Phi (0) = {2im\pi\over{g\sqrt{ L}}} \int \bar \psi \gamma_5  \psi dx . \eqno{(5.35)} $$
where $\Phi (0)$ of the boson will be discussed in detail in the next section in connection 
to the bosonization of the Schwinger model.

\subsection{Bosons of Massive Thirring Model}

The massive Thirring model has some bound states which are composed out of 
fermions and antifermions. The number of the bound states has been debated since 
the semiclassical calculation of the sine-Gordon model predicted that there should 
be many bound states in the massive Thirring model \cite{das}. On the other hand, 
the infinite momentum frame calculation and the Bogoliubov transformation method 
predict that there is only one bound state \cite{q12,q10}. This is quite reasonable since 
the interaction of the massive Thirring model is in fact the $\delta-$function 
potential which can normally possess one bound state. In addition, the Bethe ansatz 
calculations of the massive Thirring model also confirm that there is only 
one bound state in the massive Thirring model \cite{q13,b133}. In particular, 
the Bethe ansatz equations are analytically solved in the strong coupling limit 
where the semiclassical method predicts many bound states \cite{b133}.  The analytic 
solutions clearly show that there is only one bound state even in the strong coupling 
limit in this field theory model.

\subsection{Summary of Thirring Model}

In this section, we have presented a symmetry broken vacuum of the Bethe ansatz solutions 
in the Thirring model, and have shown that the true vacuum energy is 
indeed lower than the symmetric vacuum energy. 
This is quite surprising since the symmetry preserving state often 
gives the lowest energy state in quantum mechanics. However, in the field theory model, 
there is also the case in which the symmetry is spontaneously broken in the vacuum 
state, and this is indeed what is realized and observed in the Thirring model. 

In this new vacuum state, the chiral symmetry is broken, and therefore 
the momentum distribution of the negative energy state becomes similar 
to a massive fermion theory. From the distribution of the vacuum momentum, 
we can deduce the effective fermion mass. However, we should note that the fermion 
should be massless in reality, and we cannot approximate the system by the massive 
fermion field theory for a boson mass evaluation. Even though some of the physical 
observables may be calculated by the approximate scheme with the massive fermion theory, 
one should keep the massless fermion scheme in general. 

We have also calculated the one particle-one hole excitation spectrum, and 
found that the spectrum has a finite gap. From this gap energy, we can determine 
the effective fermion mass, and confirm that the effective fermion mass from the gap 
energy agrees with the one which is estimated from the vacuum momentum distribution.  

Also, we have shown that the bosonization procedure of the massless Thirring 
model has a serious defect since there is no corresponding zero mode 
of the boson field and that the massless Thirring model therefore 
cannot be fully bosonized.  

Since the massless Thirring model cannot be bosonized properly, 
there is no massless excitation spectrum in the model, and 
this is consistent with the Bethe ansatz solutions that the massless 
Thirring model has a finite gap and then the continuum spectrum starts 
right above the gap. 

Also, we should stress that the bosonization of the massless Thirring model 
has a subtlety, and one must be very careful for treating it. If one makes 
a small approximation or a subtle mistake in calculating the spectrum 
of the Hamiltonian, then one would easily obtain unphysical 
massless excitations from the massless Thirring model. 
We believe that the same care must be taken for the $SU(N)$ Thirring 
model where some approximations like the $1/N$  expansion are made and 
the massless boson is predicted \cite{wit}. When we discuss the 
large $N$ expansion, there are serious problems related to the $1/N$ 
approximation. The basic point is that they cannot take into 
account the subtlety of the dynamics. In particular, if one makes first 
the large $N$ limit, then one loses some important interactions which contribute 
to the boson mass. As Gross and Neveu pointed out in their paper \cite{q99}, 
the massless boson does not exist if they were to calculate to the higher 
orders in $1/N$. The existence of massless boson will give rise to infrared 
infinities arising from virtual states. This means that the lowest order 
approximation in $1/N$ is meaningless, and to investigate 
the infrared stability of the theory one has to work to all orders in $1/N$. 
This infra-red problems become particularly important when treating the bound state 
like boson mass.  

It is clear by now that the present results are 
in contradiction with Coleman's theorem \cite{q3}. 
Here, we have presented counter examples against Coleman's theorem, and 
the exact solutions in two dimensional field theory should correspond 
to "experimental facts". Therefore, one should figure out 
the mathematical reason why Coleman's theorem is violated in fermion 
field theory model. 
In addition to the massless Thirring model, QCD with massless fermions 
in two dimensions spontaneously breaks the chiral symmetry 
with the axial vector current conservation as we will see later. 
Therefore, the massless QCD$_2$ is also in contradiction 
with Coleman's theorem, but there is no massless boson \cite{q91}, and 
in this sense, it does not violate the theorem that there should not 
exist any massless boson in two dimensions. In reality, there is no example 
of fermion field theory models in which the symmetry of the vacuum state 
is $not$ broken due to Coleman's theorem. 
However, the basic and mathematical problem with Coleman's theorem 
is still unsolved here. But we believe that 
the basic problem of the symmetry breaking business in two dimensions must 
come from the Goldstone theorem itself for the fermion field theory as discussed 
in sections 2 and 4.


\section{ Schwinger Boson in Two Dimensional QED }

The best known model of the bosonization is the Schwinger model \cite{sch} 
which is the two dimensional QED with massless fermions. In the Schwinger 
model, the coupling constant has a mass dimension, and, due to this 
super-renormalizability, one can treat the model quite easily in many respects. 
There is no infinity at the large momentum, and therefore one does not need 
any cutoff momentum. In the Schwinger model, therefore, all the physical 
observables are described in terms of the coupling constant $g$.

\subsection{Schwinger Model}

The Schwinger model is the two dimensional QED with massless fermions. 
This is exactly solved, and this exact solution in field theory 
means that the Schwinger model can be rewritten into a new free field 
theory model. Indeed, the Schwinger model can be bosonized and becomes a free 
massive boson field theory model. This is quite interesting, but we believe 
that the Schwinger model is very special in that the interaction 
between fermions and antifermions is attractive, but it is a confining 
potential. Therefore, there exists no free fermion state, and that should 
be a strong reason why the Schwinger model becomes identical to 
the free bosonic fields.

The Lagrangian density for the Schwinger model can be written as 
$$  {\cal L} =  \bar \psi i \gamma_{\mu} D^{\mu}  \psi 
  -{1\over 4} F_{\mu\nu} F^{\mu\nu}   \eqno{ (6.1)} $$
where
$$  D_{\mu}= {\partial}_{\mu} + ig A_{\mu}, \quad
  F_{\mu\nu}= \partial_{\mu} A_{\nu} - \partial_{\nu} A_{\mu} . $$
The equation of motion for the gauge field is
$$ \partial_{\mu} F^{\mu\nu} = g j^{\nu}  \eqno{ (6.2)} $$
where the fermion current $  j_{\nu} $  
is given as
$$  j_{\nu} = :\bar \psi  \gamma_{\nu} \psi : \eqno{ (6.3)}   $$
where $ : \quad : $ denotes a normal ordering. The Dirac equation becomes 
$$  \gamma^{\mu} ({-i} \partial_{\mu} + gA_{\mu} ) 
    \psi = 0  . \eqno{(6.4)} $$
Now, we quantize the fermion field in a box with the length $L$
$$
\psi(x) = 
 \frac{1}{\sqrt{L}}\sum_{n}
\left(
      \begin{array}{c}
      a_{n} \\
      b_{n}
      \end{array}
\right)
e^{i\frac{2\pi}{L}nx}. \eqno{(6.5)} $$
Here, we take the Coulomb gauge fixing
$$ \partial_1 A^1 =0  . $$
In this case, the Hamiltonian of the Schwinger model can be written as 
$$ H= {L\over{2}} \dot{A^1}^2 +\sum_n \left({2\pi\over L} n+gA^1 
\right)a^{\dagger}_n a_n + \sum_n \left(-{2\pi\over L} n-gA^1 
\right)b^{\dagger}_n b_n  +{g^2L\over{8\pi^2}} \sum_{p \neq 0} {1\over p^2} 
\tilde{j_0}(p) \tilde{j_0}(-p) \eqno{(6.6)} $$
where $ \tilde{j_0}(p)$ denotes the momentum representation of 
the fermion current $ j_0(x)$.

The Schwinger model is solved by several methods. The bosonization is 
one of them and we will discuss it below. Also, the Schwinger model has been 
solved by the Bogoliubov transformation method. In principle, 
the Bogoliubov transformation method is an approximate scheme 
for the four fermion interaction models. 
However, the correct mass of the Schwinger boson is obtained by the Bogoliubov 
transformation method. Until now, it is not clarified 
why the Bogoliubov transformation method can give an exact mass for the Schwinger model. 
Further, the Bogoliubov transformation method reproduces the right condensate value 
of the Schwinger model which is obtained analytically. This suggest that 
the Bogoliubov vacuum state may well be a good vacuum state 
since the condensate value should exhibit some information of the vacuum structure.

\subsection{Bosonization of Schwinger Model}

In the Schwinger model, the Coulomb gauge is taken, and 
in this case, the space part of the  vector potential $A^1$ depends on 
time and corresponds to the zero mode of the boson field \cite{p2}. 
Since the fermion current $j_{\mu}$ is defined in eq.(6.3), 
the momentum representation $\tilde{j_{\mu} }$ of the current can be written 
in terms of $\rho_a (p)$ and $\rho_b (p)$ 
$$ \tilde{j_{0} }(p)= \rho_a (p)+ \rho_b (p) \eqno{(6.7a)} $$
$$ \tilde{j_{1} }(p)= \rho_a (p)- \rho_b (p) \eqno{(6.7b)} $$
where $\rho_a (p)$ and $\rho_b (p)$  are defined as 
$$ \rho_a (p) = \sum_k a^{\dagger}_{k+p} a_k \eqno{(6.8a)} $$ 
$$ \rho_b (p) = \sum_k b^{\dagger}_{k+p} b_k . \eqno{(6.8b)} $$ 
Now, we can easily prove that $\rho_a (p)$ and $\rho_b (p)$ satisfy 
the following commutation relations, 
$$ [ \rho_a (p), \rho_a (q) ] |{\rm phys} \rangle = -p\delta_{p,-q} |{\rm phys} \rangle \eqno{(6.9a)} $$
$$ [ \rho_b (p), \rho_b (q) ] |{\rm phys} \rangle = p\delta_{p,-q} |{\rm phys} \rangle . \eqno{(6.9b)} $$
These commutation relations can only be valid when these equations 
are always supposed to operate on the physical states $|{\rm phys} \rangle$. 
Here, the physical states mean that the negative energy states must be completely occupied 
if the negative energy levels are sufficiently deep. Further, in this physical state, 
there should be no particles in the positive energy states if the particle energy 
is sufficiently high. Under these conditions, eqs.(6.9) hold true as operator 
equations. 

In this case, $\tilde{j_{0} }(p)$ and $\tilde{j_{1} }(p)$ 
are related to the boson field and its conjugate field as 
$$ \tilde{j_{0} }(p) =ip\sqrt{L\over \pi} \Phi (p) \qquad {\rm for}\ \ \  p 
\not= 0 \eqno{(6.10a)} $$
$$ \tilde{j_{1} }(p) =\sqrt{L\over \pi} \Pi (p) \qquad {\rm for}\ \ \  p \not= 0  
\eqno{(6.10b)} $$
where $\Phi (p)$ and $\Pi (p)$ denote the boson field and its 
conjugate field, respectively. $L$ denotes the box length. 

It is very important to note that $\Pi (0)$ and $\Phi (0)$ are not defined in 
eqs.(6.10). In the Schwinger model, they are related to the chiral 
charge and its time derivative as 
$$ \Pi (0) = {\pi\over{g\sqrt{ L}}} Q_5 \eqno{(6.11a)} $$
$$ \Phi (0) = {\pi\over{g\sqrt{ L}}} {\dot Q}_5 \eqno{(6.11b)} $$
where ${\dot Q}_5 $ is described by the vector field $A^1$ due to 
the anomaly equation 
$$ {\dot Q}_5 ={Lg\over \pi} {\dot A}^1 . \eqno{(6.12)} $$
Here, we briefly discuss how one obtains the chiral anomaly when one regularizes 
the charge and the energy of the vacuum. The procedure to obtain eqs.(6.12) 
is shown in \cite{m1} in a clear way. 

First, we define the charges of the right and the left movers by
$$ Q_L = \sum_{n=-\infty}^{N_L} e^{\lambda ( n+ {LgA^1\over{2\pi}})} \eqno{(6.13a)} $$
$$ Q_R = \sum_{n=N_R}^{\infty} e^{\lambda ( -n- {LgA^1\over{2\pi}})} \eqno{(6.13b)} $$
where the charges are regularized in terms of the $\zeta$ function regularization. 
Here, it is important to note that one should regularize the charge with the gauge 
invariant way since the Hamiltonian has still the invariance of 
a large gauge transformation $n \rightarrow n+{LgA^1\over{2\pi}}$. 
In this case, the charge and the chiral charge of the vacuum state is defined as 
$$ Q=Q_R+Q_L \eqno{(6.14a)} $$
$$ Q_5=Q_R-Q_L . \eqno{(6.14b)} $$
The regularized charge and chiral charge become
$$ Q={2\over \lambda}+N_L+1-N_R+ O(\lambda) \eqno{(6.15a)} $$
$$ Q_5=2N_L+1+ {LgA^1\over{\pi}}. \eqno{(6.15b)} $$
Since the charge of the vacuum must be zero, we set $Q=0$ where 
we should neglect the ${2\over \lambda}$ term. From eq.(6.15b), we obtain 
the anomaly equation of eq.(6.12) by making the time derivative of $Q_5$. 
Namely, the chiral current is not conserved any more due to the anomaly.  

Further, we should regularize the vacuum energy in the same way as the charge. 
Denoting the left and right movers of the vacuum energy by $E_L^{vac}$ and $E_R^{vac}$, 
we can calculate them as
$$ E_L^{vac} ={2\pi\over L} \sum_{n=-\infty}^{N_L}\left( n+ {LgA^1\over{2\pi}}\right) 
e^{\lambda ( n+ {LgA^1\over{2\pi}})} \eqno{(6.16a)} $$
$$ E_R^{vac} ={2\pi\over L} \sum_{n=N_R}^{\infty} \left(- n- {LgA^1\over{2\pi}}\right) 
e^{\lambda (- n- {LgA^1\over{2\pi}})} . \eqno{(6.16b)} $$
Making use of eqs.(6.13), we obtain
$$ E_L^{vac} ={\pi\over{ L}} Q_L^2 \eqno{(6.17a)} $$
$$ E_R^{vac} ={\pi\over{ L}} Q_R^2 . \eqno{(6.17b)} $$
Therefore, the total vacuum energy can be written as 
$$ E^{vac} = {\pi\over{ 2L}} Q_5^2 . \eqno{(6.18)} $$
Thus, the vacuum energy part of the Hamiltonian eq.(6.6) can be written as 
$$ H^{vac} = {\pi^2\over{2g^2 L}} {\dot Q}_5^2 + 
{\pi\over{2 L}} Q_5^2 . \eqno{(6.19)} $$
If we identify the boson field $\Phi (0) $, $\Pi (0) $ as
$$ \Phi (0) ={\pi\over{g \sqrt{L}}}  Q_5  \eqno{(6.20a)} $$
$$ \Pi (0) ={\pi\over{g \sqrt{L}}}  {\dot Q}_5  \eqno{(6.20b)} $$
then we can write the vacuum part of the Hamiltonian
$$ H^{vac}= {1\over 2} {\Pi}^{\dagger} (0) \Pi (0)
+ {g^2\over{2 \pi}} {\Phi}^{\dagger} (0) \Phi (0)  . 
\eqno{(6.21)} $$

Further, the positive energy part of the kinetic energy Hamiltonian 
can be rewritten in terms of the kinetic energy of the boson Hamiltonian. 
$$  \sum_p {2\pi p\over L}  a^{\dagger}_p a_p - \sum_p {2\pi p\over L}  
b^{\dagger}_p b_p 
= {1\over 2} \sum_{p \not= 0}
\left\{  {\Pi}^{\dagger} (p) \Pi (p)
+ \left({2\pi p\over L}\right)^2 {\Phi}^\dagger (p)\Phi (p) \right\} . \eqno{(6.22)} $$
Here, we should note that the identification of the kinetic energies between 
the fermion and boson fields can be considered as operator equations with the condition 
that all the operations should be done onto physical states in fermion Fock space which 
are explained above. In this case, eq.(6.22) holds true as operator 
equations under these conditions.

Together with the Coulomb interaction part, we can write down the Hamiltonian 
for the Schwinger model as
$$ H= \sum_{p} \left\{ {1\over 2} {\Pi}^{\dagger} (p) \Pi (p)
+{1\over 2}\left({2\pi p\over L}\right)^2 {\Phi}^\dagger (p)\Phi (p) + 
{g^2\over{2 \pi}} {\Phi}^{\dagger} (p) \Phi (p) \right\}. 
\eqno{(6.23)} $$
This is just the free massive boson Hamiltonian. 

It should be important to note that the Schwinger model has the right 
zero mode in the Hamiltonian of the boson field. However, as we saw 
in section 5, there is no corresponding zero mode in the massless Thirring model, 
and this leads to the finite gap of the spectrum in the massless Thirring 
model, which is indeed consistent with the fact that there should exist no physical massless 
boson in two dimensions. 

\subsection{QED$_2$ with Massive Fermions}

It should be interesting to make some comments on the boson spectrum 
in QED$_2$ with massive fermions. The Lagrangian density is just the same 
as eq.(6.1) with the fermion mass term
$$  {\cal L} =  \bar \psi (i \gamma_{\mu} D^{\mu}-m_0 ) \psi 
  -{1\over 4} F_{\mu\nu} F^{\mu\nu}  . \eqno{ (6.24)} $$
At the massless limit, there is only one boson which is the Schwinger boson. 
For the finite but small fermion mass $m_0$ region, the lowest boson mass can be written 
as \cite{berg}
$$  {\cal M} \simeq {g\over{\sqrt{\pi}}} + e^{\gamma} m_0  \eqno{ (6.25)} $$
where $\gamma$ denotes Euler's constant. 

As the mass increases, the number of the bosonic bound states increases \cite{q11}. 
When the mass is much larger than the coupling constant ${g\over{\sqrt{\pi}}} $, 
then the system becomes just the same as the nonrelativistic quantum mechanics. 
The particle and antiparticle are interacting with each other through the linear 
rising potential. In this case, the vacuum state is just like the perturbative 
vacuum, and there is no fermion condensate.


\begin{figure}[htbp]
\centering
\includegraphics[width=0.7\textwidth]{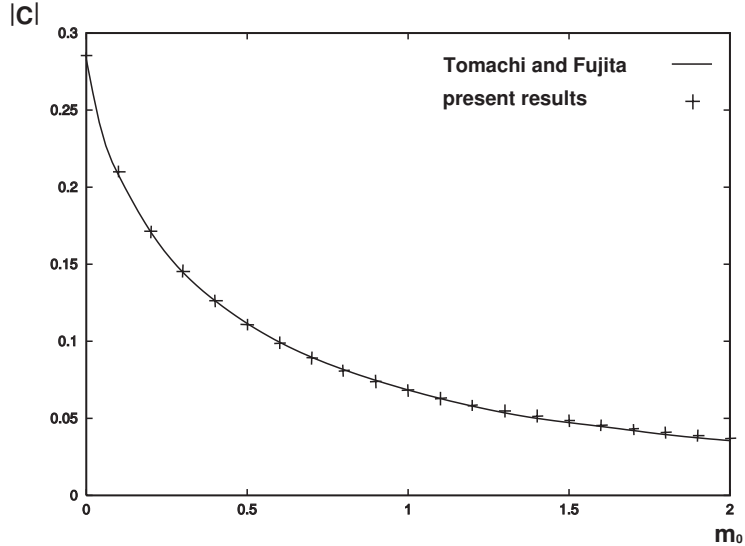}
\caption{The absolute values of the condensate for massive QED$_2$ 
are plotted as the function of the fermion mass $m_0$ in the 
small mass regions. The solid line is calculated in \cite{q11} 
while the crosses are evaluated in the present paper with the massless 
fermion basis.}
\label{fig3}
\end{figure}

In this respect, the interesting region must be the one in which 
the fermion mass is much smaller then the coupling constant ${g\over{\sqrt{\pi}}} $, 
but it is still finite. If the fermion mass is finite, then there is no chiral 
symmetry in the Lagrangian density. This means that there cannot be 
any symmetry breaking phenomena in the massive fermion QED$_2$. 
However, if one evaluates the chiral condensates, then one finds a finite 
condensate value for the massive fermion case. In other words, the fermion 
condensate value is a smooth function of the fermion mass $m_0$ \cite{q11} 
as we show in Fig. 3.

The chiral condensate value for the massive fermion QED$_2$ can be written 
$$ \langle \Omega | {1\over L} \int \bar \psi  \psi dx |\Omega \rangle 
\simeq {g\over{\sqrt{\pi}}} {e^\gamma \over{2\pi}} + O(m_0) \eqno{ (6.26)} $$
This strongly suggests that the chiral condensate value is not the consequence 
of the symmetry breaking, but it indicates the vacuum structure 
how many of the virtual particle pairs can be found in the vacuum. Or in other words, 
the chiral condensate is related to the change of the momentum distributions of 
the negative energy particles in comparison with the symmetric distributions 
of the free particles in the vacuum state.


\section{ Bosons in Two Dimensional QCD }

QCD in two dimensions presents a good example of the strong interaction 
models since it can be solved to a good accuracy. The basic structure of QCD$_2$ 
is similar to QED$_2$ in that both of the models are super-renormalizable. 
In addition, there is no transverse field, and therefore it becomes identical 
to the four fermion interaction model in the field theory where the interaction 
is described in terms of the fermion fields only. 

The main difference between QCD$_2$ and QED$_2$ is of course due to the color 
degrees of freedom. Concerning the symmetry breaking phenomena, there is no anomaly 
in QCD$_2$ since there is no anomaly term which has a color singlet state in two dimensions. 
Therefore, the axial vector current is conserved after the chiral symmetry is 
broken in contrast to QED$_2$. In this respect, it should be quite 
interesting to study whether the vacuum state of QCD$_2$ breaks the chiral 
symmetry or not. Here, we show that the chiral symmetry is spontaneously 
broken in QCD$_2$, and the chiral condensate value is finite. However, 
there is no massless boson and the boson spectrum is just similar to that 
of QED$_2$. 

The boson mass spectrum in QCD$_2$ has been extensively studied 
by the light cone method \cite{q01,q022,q02}. 
In particular, QCD$_2$ with the $1/N_c$ expansion proposed by 't Hooft has 
presented  interesting results on the boson mass 
spectrum \cite{q21,q210,q211,q22,q23}. The boson mass 
vanishes when the fermion mass becomes zero. However, this is not allowed since 
the massless boson cannot {\it physically} exist 
in two dimensional field theory \cite{q3,q4}. 
Unfortunately, this problem of the puzzle has never been seriously considered 
until now, apart from unrealistic physical pictures. People believe that 
the large $N_c$ limit is special because one takes $N_c$ infinity. 
But the infinity in physics means simply that the $N_c$ must be 
sufficiently large, and, in fact, as we show below, physical observables 
at $N_c=50$ are just the same as those of $N_c = \infty$. 

Further, this boson spectrum of large $N_c$ QCD$_2$ was confirmed 
by the light cone calculations 
with $SU(2)$ and $SU(3)$ colors \cite{q022}. 
Indeed, the mass of the boson in the light cone calculations 
is consistent with the 't Hooft spectrum of the boson even though the latter is 
evaluated by the $1/N_c$ approximation. However, the fact that the light cone calculation 
predicts massless bosons is rather serious since 
the light cone calculation for $SU(2)$ does not seem to make any unrealistic 
approximations, apart from the trivial vacuum. 

However, there is an interesting indication that the light cone vacuum is 
not trivial, and indeed there is a finite condensate even for the large 
$N_c$  QCD$_2$ \cite{q05,q06,q03,q04}. What does this mean ? 
This suggests that one has to consider 
the effect of the complicated vacuum structure for the boson mass as long as 
one calculates the boson mass  with Fock space expansions. On the other hand,  
the calculation for the boson spectrum by 't Hooft is based on the trivial vacuum,  
but, instead he could sum up all of the intermediate 
fluctuations of the fermion and antifermion pairs. This should be 
equivalent to considering the true vacuum structure in the Fock space basis. 
That is, the same spectrum of bosons must be obtained both by 
the Fock space expansion with the true vacuum and by the sum of 
all the Feynman diagrams with the trivial vacuum if they are 
treated properly. 

For this argument, people may claim that QED$_2$ is exactly described by the 
naive light cone calculation with the trivial vacuum, and therefore, QCD$_2$ 
may well be treated just in the same way as the QED$_2$ case. However, 
one may well have some uneasy feeling for the fact that the naive 
light cone calculation cannot reproduce the condensate value of QED$_2$. 

In this section, we show that the light cone calculation based 
on the Fock space expansion with the trivial 
vacuum is not valid for QCD$_2$. One has to consider properly the effect 
of the complicated vacuum structure. Here, we present the calculation with 
the Bogoliubov vacuum in the rest frame, 
and show that the present calculation reproduces the right condensate values. 
Indeed, we can compare the present results with the condensate value  
as predicted by the $1/N_c$ expansion\cite{q05,q06,q03,q04}, 
$$ C_{N_c}=-{N_c\over{\sqrt{12}}} \sqrt{N_cg^2\over{2\pi}} . \eqno{(7.1)} $$ 
The present calculation of the condensate value for the $SU(2)$ color 
is $C_2=-0.495$  ${g\over{\sqrt{\pi}}} $ which should be 
compared with the $-0.577$  ${g\over{\sqrt{\pi}}} $ from the $1/N_c$ expansion, and 
$C_3=-0.995$  ${g\over{\sqrt{\pi}}} $ for the $SU(3)$ color  compared 
with $-1.06$  ${g\over{\sqrt{\pi}}} $ of the $1/N_c$ expansion. 
For the larger value of $N_c$ (up to $N_c=50$), we obtain the condensate values 
which perfectly agree with the prediction of $C_{N_c}$ in eq.(7.1). 

Further, we show that the boson masses for QCD$_2$ 
with $SU(2)$ and $SU(3)$ colors are finite 
even though the fermion mass is set to zero. 
In fact, the boson mass is found 
to be ${\cal M}_2=0.467$ ${g\over{\sqrt{\pi}}} $ for the $SU(2)$, 
and ${\cal M}_3=0.625$ ${g\over{\sqrt{\pi}}} $ 
for the $SU(3)$ color for the massless fermions. 
Further, the present calculations of the boson mass up to $N_c=50$ 
suggest that the boson mass ${\cal M}_{N_c}$ for $SU(N_c)$ can be described 
for the large $N_c$ by the following phenomenological expression at 
the massless fermion limit, 
$$  {\cal M}_{N_c}={2\over 3}\sqrt{{N_cg^2\over{3\pi}}} . \eqno{(7.2)} $$
Also, we calculate the boson mass at the large $N_c$ with the finite 
fermion mass. From the present calculations, we can express the boson mass 
in terms of the phenomenological formula for the small fermion mass $m_0$ region, 
$$  {\cal M}_{N_c} \approx \left( {2\over 3}{\sqrt{2\over 3}}
+{10\over{3}}{m_0\over{\sqrt{N_c}}} \right) 
\sqrt{{N_cg^2\over{2\pi}}}  \eqno{(7.3)} $$
where $m_0$ is measured in units of ${g\over{\sqrt{\pi}}}$. 

The above expression (eq.(7.3)) can be compared with the calculation 
by Li et al. \cite{q210} who employed the $1/N_c$ expansion of 't Hooft model 
in the rest frame \cite{q211}. 
It turns out that their calculated boson mass for their smallest fermion 
mass case is consistent with the above equation, though their calculated 
values are slightly smaller than the present results. 
   
In addition, we examine the validity of the light cone calculation 
for QED$_2$. It is shown that the boson mass for the QED$_2$ case 
happens to be not very sensitive to the condensate value, and that 
the spectrum can be reproduced by the light cone calculations 
with the trivial vacuum as well as with the condensate value only 
with positive momenta. 
Therefore, we believe that the QED$_2$ case is accidentally reproduced 
by the light cone calculation with the trivial vacuum state even though 
we do not fully understand why this accidental agreement occurs. 
On the other hand, the QCD$_2$ case is quite different. 
The boson mass calculated with the trivial 
vacuum is zero at the massless fermion limit. Further, the calculation 
in the light cone with the condensate value only with the positive momenta 
are not stable against the infra-red singularity of the light cone equations. 

The present calculations are based on the Fock space expansion, and, in this 
calculation, we only consider the fermion and anti-fermion (two fermion) 
space.  For QED$_2$, it is shown that the fermion and anti-fermion 
space reproduces the right Schwinger boson \cite{q11}. 
That is, the four fermion spaces 
do not alter the lowest boson energy in QED$_2$. However, there is 
no guarantee that there are finite effects on the lowest boson mass 
from the four fermion spaces in QCD$_2$. This point is not examined 
in this paper, and should be worked out in future. 

Here, we examine the RPA calculations and show that the boson mass 
for QED$_2$ with the RPA equations deviates from the Schwinger boson. 
That means that the agreement achieved by the Fock space expansion 
is destroyed by the RPA calculation. 
Further, we calculate the boson mass for QCD$_2$ 
with $SU(2)$ and the large $N_c$ limit. It turns out that the boson 
mass vanishes when the fermion mass is equal to the critical value 
and that it becomes imaginary when the fermion mass is smaller than 
the critical value. This is obviously unphysical at the massless fermion limit, 
and  is closely related to the fact 
that the RPA equations are not Hermitian, 
and therefore  we should examine its physical meaning in future. 

From the present calculation, we learn that the chiral symmetry in massless  QCD$_2$ 
is spontaneously broken without the anomaly term, in contrast to the Schwinger model. 
But the boson mass is finite, and therefore 
there is no Goldstone boson in this field theory model. 
Thus, the present result confirms that 
the Goldstone theorem \cite{q1,q2} does not hold for the fermion field theory 
as discussed in section 2. This indicates that the anomaly term has 
little to do with the chiral symmetry breaking. This is reasonable since 
the anomaly term arises from the conflict between the gauge invariance 
and the chiral current conservation when regularizing the vacuum, 
and this is essentially a kinematical effect. 
On the other hand, the symmetry breaking is closely related to  
the vacuum energy which of the vacuum states should have the lowest energy, 
and therefore it is the consequence of the dynamical effects in the vacuum. 

\subsection{Bogoliubov Transformation in QCD$_2$}
In this section, we discuss the Bogoliubov transformation in QCD$_2$. 
The Lagrangian density for QCD$_2$ with $SU(N_c)$ color is described as 
$$ {\cal L}= 
\bar{\psi}(i\gamma^{\mu}\partial_{\mu}-g\gamma^{\mu}A_{\mu} -m_0)\psi 
-\frac{1}{4}F^{a}_{\mu\nu}F^{a\mu\nu},
 \eqno{(7.4)} $$
where $F_{\mu\nu}$ is written as
$$ F_{\mu\nu} = \partial_{\mu}A_{\nu}-\partial_{\nu}A_{\mu}+ig[A_{\mu},A_{\nu}],
\quad A_{\mu} = A^{a}_{\mu}T^{a}, \ \ \  
T^{a} = \frac{\tau^{a}}{2} .  $$ 
Here, $m_0$ denotes the fermion mass, and at the massless limit, the Lagrangian density 
has a chiral symmetry.

Now, we first fix the gauge by 
$$ A^a_1 =0 . $$  
This gauge fixing has been employed by most of the calculations which have been 
done up to now. This gauge is simple but cannot describe the zero mode 
even though the spectrum is properly described by this gauge. 
In this gauge, the equation of motion for $A^a_0$ becomes
$$\partial_{1}^{2}A^a_0 = -gj^a_0,  \eqno{(7.5)} $$ 
where $j^a_0$ is the fermion current defined by
$$j^{a}_0 = \psi^\dagger\frac{\tau^{a}}{2}\psi.   $$
The Hamiltonian can be written as
$$ H = \int dx\left[ -i{\bar \psi}\gamma^{1}\partial_{1}\psi +m_0\bar{\psi}\psi 
+\frac{g}{2}j^{a}_0A^{a}_0 \right].\eqno{(7.6a)} $$
Now, we quantize the Hamiltonian  of QCD$_2$ with $SU(N_c)$ color, and 
it can be written as 
$$ H = \sum_{n,\alpha}p_{n}
\left(a_{n,\alpha}^{\dagger}a_{n,\alpha}-b_{n,\alpha}^{\dagger}b_{n,\alpha}
\right)  
 +m_0 \sum_{n,\alpha}
\left(a_{n,\alpha}^{\dagger}b_{n,\alpha}+b_{n,\alpha}^{\dagger}a_{n,\alpha}
\right)  $$
$$ -\frac{g^2}{4N_cL}\sum_{n,\alpha,\beta}\frac{1}{p^{2}_{n}}
\left({\tilde j}_{1,n,\alpha\alpha}+{\tilde j}_{2,n,\alpha\alpha}
\right)
\left({\tilde j}_{1,-n,\beta\beta}+{\tilde j}_{2,-n,\beta\beta}
\right)
\nonumber $$
$$ +\frac{g^2}{4L}\sum_{n,\alpha,\beta}\frac{1}{p^{2}_{n}}
\left({\tilde j}_{1,n,\alpha\beta}+{\tilde j}_{2,n,\alpha\beta}
\right)
\left({\tilde j}_{1,-n,\beta\alpha}+{\tilde j}_{2,-n,\beta\alpha}
\right), \eqno{(7.6b)} $$
where
$$ 
{\tilde j}_{1,n,\alpha\beta} = 
\sum_{m}a_{m,\alpha}^{\dagger}a_{m+n,\beta} \eqno{(7.7a)} $$
$$ {\tilde j}_{2,n,\alpha\beta} = 
\sum_{m}b_{m,\alpha}^{\dagger}b_{m+n,\beta} . \eqno{(7.7b)} $$
Now, we define new fermion operators by the Bogoliubov transformation, 
$$ a_{n,\alpha}=\cos\theta_{n,\alpha}c_{n,\alpha}+\sin\theta_{n,\alpha}
d_{-n,\alpha}^{\dagger} \eqno{(7.8a)} $$
$$
b_{n,\alpha}=-\sin\theta_{n,\alpha}c_{n,\alpha}+\cos\theta_{n,\alpha}
d_{-n,\alpha}^{\dagger} \eqno{(7.8b)} $$
where $ \theta_{n,\alpha}$ denotes the Bogoliubov angle. 

In this case, the Hamiltonian of QCD$_2$ can be written as 
$$ H = \sum_{n,\alpha}E_{n,\alpha}(c^\dagger_{n,\alpha}c_{n,\alpha}
+d^\dagger_{-n,\alpha}d_{-n,\alpha}) +H' \eqno{(7.9)} $$
where
\begin{align*}
 E^2_{n,\alpha} = & \left\{p_{n}+\frac{g^2}{4N_cL}\sum_{m,\beta}
\frac{(N_c\cos 2\theta_{m,\beta}-\cos 2\theta_{m,\alpha})}{(p_m-p_n)^2} \right\}^2 \\
& +\left\{m_0+ \frac{g^2}{4N_cL}\sum_{m,\beta}{
(N_c\sin 2\theta_{m,\beta}-\sin 2\theta_{m,\alpha}) 
\over{{(p_m-p_n)^2} }} \right\}^2 .
 \tag{7.10}
\end{align*}
$H'$ denotes the interaction Hamiltonian in terms of the new operators but 
is quite complicated, and therefore it is not given here. 

The energy of the single particle state [eq.(7.10)] does not have 
a proper energy and momentum dispersion relation. This means that 
the quasi-particle states cannot be a physical state. However, this 
is quite reasonable since fermions in QCD$_2$ are confined and 
they can never be observed. Physical observables are bosonic states 
which are to be evaluated below. 
 
The conditions that the vacuum energy is minimized give the constraint 
equations which can determine the Bogoliubov angles 
$$ \tan 2\theta_{n,\alpha} =
\frac{m_0+\frac{g^2}{4N_cL}\sum_{m,\beta}
\frac{(N_c\sin 2\theta_{m,\beta}-\sin 2\theta_{m,\alpha})}
{(p_m-p_n)^2}}
{p_{n}+\frac{g^2}{4N_cL}\sum_{m,\beta}\frac{(N_c\cos 2\theta_{m,\beta}-
\cos 2\theta_{m,\alpha})}{(p_m-p_n)^2}} . \eqno{(7.11)} $$
The conditions of eq.(7.11) can be also derived from the requirement that 
the $(c_{n,\alpha}d_{n,\alpha}+h.c.) $ terms should vanish. 

In this case, the condensate value $C_{N_c}$ is written as 
$$ C_{N_c} ={1\over L} \sum_{n,\alpha} \sin 2\theta_{n,\alpha} . \eqno{(7.12)} $$
Now, we can calculate the boson mass for the $SU(N_c)$ color. First,  we 
define the wave function for the color singlet boson as 
$$ |\Psi_{K}\rangle = \frac{1}{\sqrt{N_c}}\sum_{n,\alpha}f_{n}
c^\dagger_{n,\alpha}d^\dagger_{K-n,\alpha}|0\rangle .  \eqno{(7.13)} $$
In this case, the boson mass can be described as
\begin{align*}
 {\cal M} = & \; \langle \Psi_{K}|H|\Psi_{K}\rangle \\
     = & \;\frac{1}{N_c}\sum_{n,\alpha}\left(E_{n,\alpha} + E_{n-K,\alpha}\right)|f_n|^2 \\
&+\frac{g^2}{2N_c^2L}\sum_{l,m,\alpha}\frac{f_{l}f_{m}}{(p_l-p_m)^2}
  \cos(\theta_{l,\alpha}-\theta_{m,\alpha})
  \cos(\theta_{l-K,\alpha}-\theta_{m-K,\alpha}) \\
&-\frac{g^2}{2N_cL}\sum_{l,m,\alpha,\beta}\frac{f_{l}f_{m}}{(p_l-p_m)^2}
  \cos(\theta_{l,\alpha}-\theta_{m,\beta})
  \cos(\theta_{l-K,\alpha}-\theta_{m-K,\beta}) \\
& +\frac{g^2}{2N_c^2L} \sum_{l,m,\alpha,\beta}\frac{f_{l}f_{m}}{K^2}
  \sin(\theta_{l-K,\alpha}-\theta_{l,\alpha})
  \sin(\theta_{m,\beta}-\theta_{m-K,\beta}) \\
& -\frac{g^2}{2N_cL} \sum_{l,m,\alpha}\frac{f_{l}f_{m}}{K^2}
  \sin(\theta_{l-K,\alpha}-\theta_{l,\alpha})
  \sin(\theta_{m,\alpha}-\theta_{m-K,\alpha}) . \tag{7.14} 
\end{align*}
$$$$
This equation can be easily diagonalized together with the Bogoliubov 
angles, and we obtain the boson mass. Here, we note that the treatment 
of the last two terms should be carefully estimated since the apparent 
divergence at $K=0$ is well defined and finite.

\subsection{  Condensate and Boson Mass in $SU(2)$ and $SU(3)$ }

Here, we present the calculated results of the condensate values 
and the boson mass in QCD$_2$ with the $SU(2)$ and $SU(3)$ colors. 
Table 3 shows the condensate and the boson mass for the two 
different vacuum states, one with the trivial vacuum and 
the other with the Bogoliubov vacuum. As can be seen, the 
condensate values for the $SU(2)$ and $SU(3)$ are already close 
to the predictions  by the $1/N_c$ expansion of eq.(7.1) \cite{q05,q06,q04}. 
The boson masses for the $SU(2)$ and $SU(3)$  are, for the first time, 
obtained as the finite value. Unfortunately, we cannot compare our 
results with any other predictions. But compared with the Schwinger 
boson, the boson masses for the $SU(2)$ and $SU(3)$ are in the same order 
of magnitude. 

In Fig. 4, we present the fermion mass dependence of the condensate 
values for the $SU(2)$ and $SU(3)$ cases. As can be seen, the condensate 
becomes a finite value at the massless limit. It decreases as the function of 
the fermion mass $m_0$. This tendency is just the same as the condensate of 
QED$_2$ \cite{q11,hos}. 

Also, in Fig. 5, we show   
the calculated results of the boson mass as the function of the $m_0$ 
for $SU(2)$ and $SU(3)$. At the massless limit, the boson mass 
becomes a finite value, and the $m_0$ dependence is linear. 
This is exactly the same as the QED$_2$ case \cite{q11,q15}. 


%

\begin{figure}[htbp]
\centering
\includegraphics[width=0.7\textwidth]{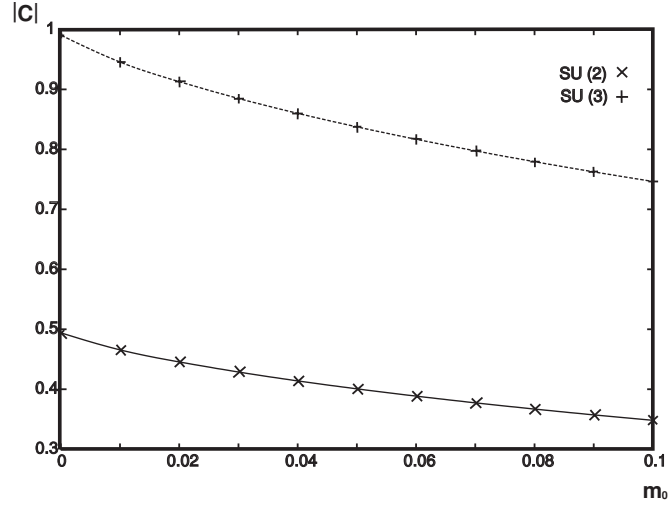}
\caption{The absolute values of the condensate for $SU(2)$ and $SU(3)$ colors 
are plotted as the function 
of the fermion mass $m_0$ in the very small mass regions. 
The solid and dashed lines are shown to guide the eyes.}
\label{fig4}
\end{figure}

%

\begin{figure}[htbp]
\centering
\includegraphics[width=0.7\textwidth]{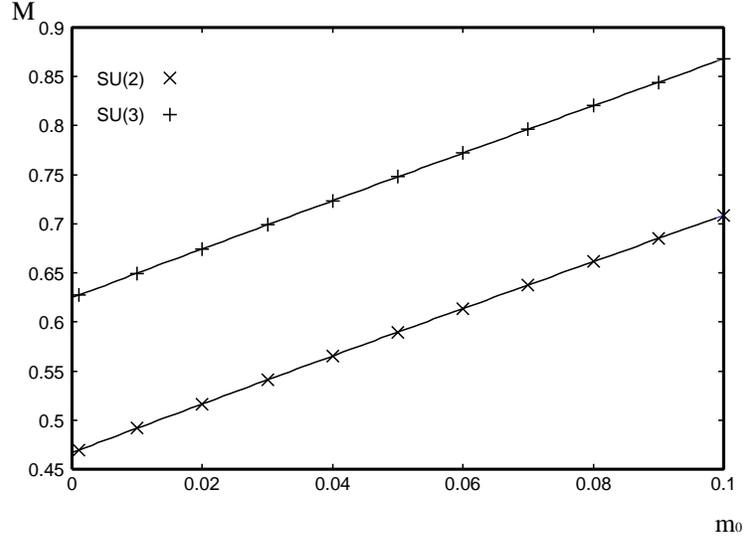}
\caption{The boson masses for $SU(2)$ and $SU(3)$ colors are plotted as the function 
of the fermion mass $m_0$ in the very small mass regions. The solid lines 
are shown to guide the eyes.}
\label{fig5}
\end{figure}


The present calculations show that both of the 
values (condensate and boson mass) are  a smooth function 
of the fermion mass $m_0$. This means 
that the vacuum structure has no singularity at the massless limit. 
This must be due to the fact that the coupling constant $g$ has 
the mass dimension and therefore, physical quantities are expressed 
by the coupling constant $g$ even at the massless limit 
of the fermion. This is in contrast to the Thirring model 
where the massless limit is a singular point. 
In the Thirring model, the coupling constant has no dimension, and 
therefore, at the massless limit, physical quantities 
must be described by the cutoff $\Lambda$.




\begin{table}[h]
\caption{We show the condensate value $C_{N_c}$ and the boson mass ${\cal M}$ of  $SU(2)$ 
and $SU(3)$ QCD$_2$ in rest frame
in units of \  $g/\sqrt{\pi}$ with $m_0=0$}
\begin{center}
\vspace{0.2cm}
\begin{tabular}{|c||c|c|c|}
\hline
\ $SU(2)$ & Trivial  & Bogoliubov  & $1/N_c$  \rule[-2mm]{0pt}{6mm}\\
\hline
\hline
$C_2$ & 0 & $-0.495$ & $-0.577$  \\
\hline
${\cal M}$ & $-\infty$    & 0.467  & 0  \\
\hline
\end{tabular}
\begin{tabular}{|c||c|c|c|}
\hline
\ $SU(3)$ & Trivial  & Bogoliubov  & $1/N_c$   \rule[-2mm]{0pt}{6mm}\\
\hline
\hline
$C_3$ & 0 & $-0.995$ & $-1.06$  \\
\hline
${\cal M}$ & $-\infty$   & 0.625  & 0  \\
\hline
\end{tabular} 
\end{center}
\end{table}

In Table 3, we show the condensate values and the boson mass 
of QCD$_2$ in the rest frame.  
Here, the minus infinity of the boson mass in the trivial vacuum 
is due to the mass singularity $ \ln (m_0) $ as explained in ref. \cite{q11}


\subsection{  Condensate and Boson Mass in $SU(N_c)$ }

Here, we  carry out the calculations of the condensate 
and the boson mass for the large $N_c$ values of $SU(N_c)$  
up to $N_c=50$. In Fig. 6, we show the calculated condensate values 
(denoted by crosses) as the function 
of the $N_c$ together with the prediction of the $1/N_c$ expansion 
as given in eq.(7.1). As can be seen, the calculated condensate values 
agree very well 
with the prediction of the $1/N_c$ expansion if the $N_c$ is larger than 10. 
Further, the calculated  boson masses (denoted by crosses) 
are shown in Fig. 7 as the function of $N_c$. 
It is found that they can be described by the following formula [eq.(7.2)] 
for the large $N_c$ values, 
$$  {\cal M}_{N_c}={2\over 3}\sqrt{{N_cg^2\over{3\pi}}} . \eqno{(7.2)} $$ 
Indeed, the calculated boson masses for $N_c $ larger than 
$N_c =10 $ perfectly agree with the predicted value of eq.(7.2).

%

\begin{figure}[htbp]
\centering
\includegraphics[width=0.7\textwidth]{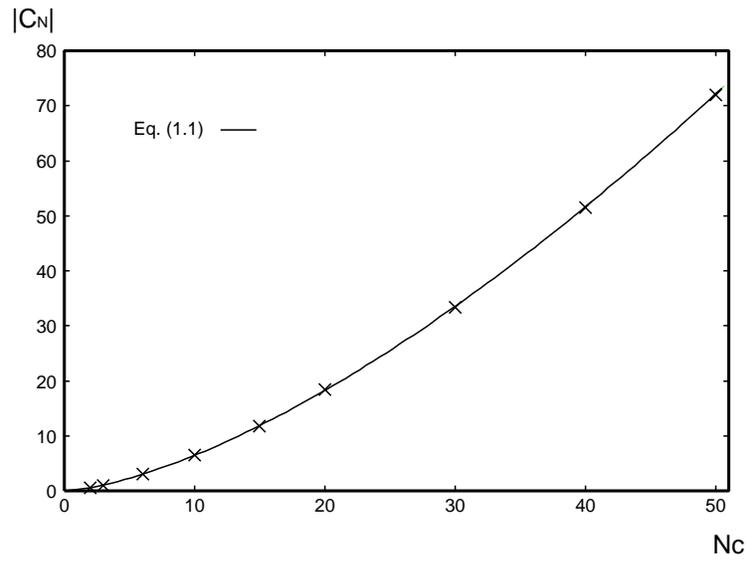}
\caption{The absolute values of the condensate for $SU(N_c)$ colors 
are plotted as the function of $N_c$. The crosses are the calculated 
values while the solid line is the prediction of eq.(7.1).}
\label{fig6}
\end{figure}


\begin{figure}[htbp]
\centering
\includegraphics[width=0.7\textwidth]{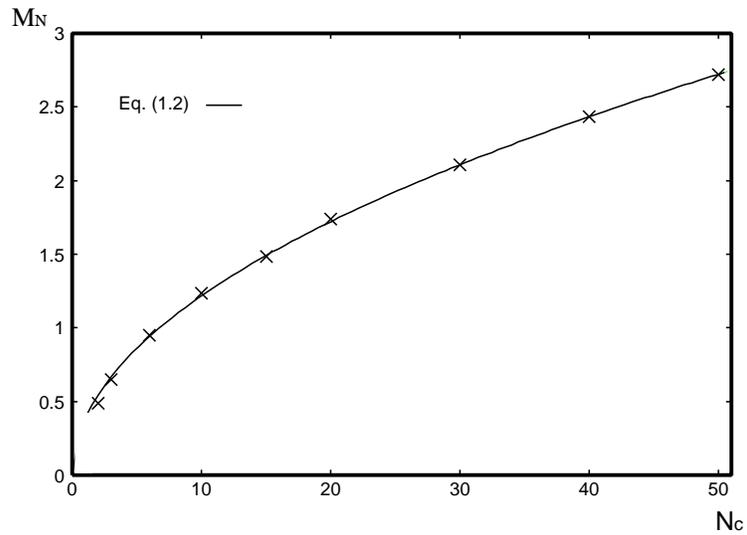}
\caption{The boson masses for $SU(N_c)$ colors with the massless fermion 
are plotted as the function of $N_c$. The crosses are the calculated 
values while the solid line is the prediction of eq.(7.2). }
\label{fig7}
\end{figure}

The present calculations show that the second excited state for $SU(N_c)$ 
colors is higher than the twice of the boson mass at the massless fermions. 
Therefore, there is only one bound state in QCD$_2$ with the $SU(N_c)$. 
This indicates that eq.(7.2) must be the full boson spectrum 
for QCD$_2$ with massless fermions. This indicates that QCD$_2$ with 
massless fermions may well be bosonized like the Schwinger model since 
the two bosons cannot make any bound states. 

Now, we present the calculations of the boson mass for the finite fermion mass 
$m_0$ cases. Here, we limit ourselves to the $m_0$ (in units of ${g\over{\sqrt{\pi}}}$) 
which is smaller than unity. 
In Fig. 8, we show the calculated  values of the boson mass 
as the function of  ${m_0\over{\sqrt{N_c}}}$ for several cases of the fermion 
mass $ m_0 $ and the color $N_c$. The present calculation is 
carried out up to the $N_c=50$ case which is sufficiently large enough 
for the large $N_c$ limit of the 't Hooft model. 
The solid line in Fig. 8 is obtained as the following phenomenological 
formula of the fit to the numerical data 
$$  {\cal M}_{N_c} \approx \left( {2\over 3}{\sqrt{2\over 3}}
+{10\over{3}}{m_0\over{\sqrt{N_c}}} \right) 
\sqrt{{N_cg^2\over{2\pi}}} . \eqno{(7.3)} $$




\begin{figure}[htbp]
\centering
\includegraphics[width=0.7\textwidth]{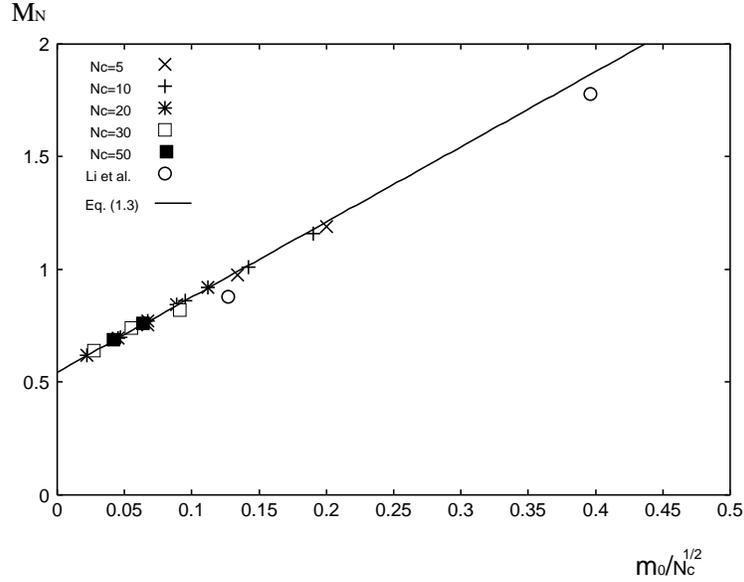}
\caption{The boson masses in units of $\sqrt{{N_cg^2\over{2\pi}}}$ 
for $SU(N_c)$ colors with the massive fermion are plotted 
as the function of $m_0/\sqrt{N_c}$. The crosses, 
circles and squares are the calculated 
values while the solid line is the prediction of eq.(7.3).}
\label{fig8}
\end{figure}

Now, we want to compare the present results with the old calculations 
by Li et al. \cite{q21,q210} who obtained the boson mass by solving the 't Hooft equations 
for QCD$_2$ with the large $N_c$ limit in the rest frame. 
Li et al. obtained the boson mass for their smallest fermion mass 
of $ m_0=0.18 \sqrt{{N_c\over{2}}} $
$$  {\cal M}_{\infty} = 0.88 \sqrt{{N_cg^2\over{2\pi}}} . \eqno{(7.15)} $$
There are also a few more points of their calculations with larger fermion 
mass cases. In Fig. 8, we plot the boson masses calculated by Li et al. by 
the white circles which should be compared with the solid line. 
As can be seen, the boson mass obtained by 
Li et al. is close to the present calculation. It should be noted that 
their calculations were carried out with rather small number 
of the basis functions in the numerical evaluation, and therefore, the 
accuracy of their calculations may not be very high, in particular, for 
the small fermion mass regions.  

Unfortunately, however, Li et al. made a wrong conclusion 
on the massless fermion limit 
since their calculated point of $ m_0=0.18 \sqrt{{N_c\over{2}}} $  
was the smallest fermion mass. Obviously, this value of the fermion mass 
was by far too large to draw any conclusions on the massless fermion limit.

\subsection{    QCD$_2$ in Light Cone }

HereCwe evaluate the boson mass in the light cone. For this, we follow 
the prescription in terms of the infinite momentum frame \cite{q12,q15} 
since this has a good connection to the rest frame calculation. 
In this frame, we can calculate the boson mass with and without 
the condensate in the light cone. But in evaluating the condensate, 
we only consider the positive momenta. 
The equation for the boson mass square for the $SU(2)$ case becomes
\begin{align*}
{\cal M}^2  = & \; m^2_{0}\int dxf(x)^2  \biggl(\frac{1}{x} +\frac{1}{1-x}\biggr)   \\
& +\frac{3g^2}{16\pi}\int dxdy \frac{f(x)^2}{(x-y)^2}
   \bigl( \cos 2\theta_{y,1} +\cos 2\theta_{1-y,1}  
   +\cos 2\theta_{y,2}+\cos 2\theta_{1-y,2} \bigr) \\
&-\frac{g^2}{4\pi}\int dxdy\frac{f(x)f(y)}{(x-y)^2}
  \biggl[ \;
  \frac{1}{2}\cos(\theta_{x,1}-\theta_{y,1})\cos(\theta_{x-1,1}-\theta_{y-1,1}) \\
& \hspace{3.5cm} +\frac{1}{2}\cos(\theta_{x,2}-\theta_{y,2}) \cos(\theta_{x-1,2}-\theta_{y-1,2}) \\
& \hspace{3.5cm} +\cos(\theta_{x,1}-\theta_{y,2})\cos(\theta_{x-1,1}-\theta_{y-1,2}) \\
& \hspace{3.5cm} +\cos(\theta_{x,2}-\theta_{y,1})\cos(\theta_{x-1,2}-\theta_{y-1,1}) \biggr] \\
&-\frac{g^2}{8\pi}\int dxdyf(x)f(y) \biggl[ \; \sin(\theta_{x,1}-\theta_{x-1,1})
     \sin(\theta_{y-1,1}-\theta_{y,1}) \\
& \hspace{3.5cm} +\sin(\theta_{x,2}-\theta_{x-1,2}) \sin(\theta_{y-1,2}-\theta_{y,2}) \\
& \hspace{3.5cm} -\sin(\theta_{x,1}-\theta_{x-1,1}) \sin(\theta_{y-1,2}-\theta_{y,2}) \\
& \hspace{3.5cm} -\sin(\theta_{x,2}-\theta_{x-1,2}) \sin(\theta_{y-1,1}-\theta_{y,1}) \biggr]
 \tag{7.16} 
\end{align*}
Here, all of the momenta are positive. 
This can be easily evaluated, and we obtain the condensate 
values and the boson mass as given in Table 4. 
We note here that both of the values become smaller 
as the function of the fermion mass, and finally they 
vanish to zero. This is exactly what is observed in the light 
cone calculations. Since the light cone calculations cannot reproduce 
the condensate values which are finite as predicted in ref. \cite{q03}, 
the light cone calculations must have some problems. 
In Table 4, we also show the calculations of the infinite momentum frame 
with the positive momenta only. However, the numerical calculations are 
not stable against the infra-red singularity of the light cone. 
At the present stage, we do not know how to evaluate them properly, and 
we do not fully understand what is wrong with the light cone. 


\begin{table}


\caption{We show the condensate value $C_{N_c}$ and the boson mass ${\cal M}$ of 
$SU(2)$ QCD$_2$in infinite momentum frame in units of \  $g/\sqrt{\pi}$  with $m_0=0$. 
Here, $**$ indicates that there is no stable solution.}

\begin{center}
\vspace{0.2cm}
\begin{tabular}{|c||c|c|c|}
\hline
\ $SU(2)$ & Trivial  & Bogoliubov ($p>0$)  & $1/N_c$ \rule[-2mm]{0pt}{6mm} \\
\hline
\hline
$C_2$ & 0 & $**$  & $-0.577$  \\
\hline
${\cal M}$ & $0$   & $**$  & 0  \\
\hline
\end{tabular} 
\end{center}

\end{table}

\subsection{  Examination of  't Hooft Model }

Here, we discuss the boson mass of QCD$_2$ with $SU(N_c)$ color 
in the large $N_c$ limit. 
This model is solved by 't Hooft who sums up all of the Feynman diagrams 
in the $1/N_c$ expansion and obtains the equations for the boson mass. 
In principle, the 't Hooft equations must be exact up to the order of $1/N_c$. 
Therefore, one does not have to consider the effect of the vacuum since 
the 't Hooft equations take into account all of the fluctuations of the intermediate 
fermion and antifermion pairs. 
Therefore, it is expected that the right boson mass can be obtained 
from the equations at the order of $1/N_c$. 

The present calculations of the boson mass with the $SU(N_c)$ colors 
show that the boson mass can be well described by 
$ {\cal M}_{N_c}={2\over 3}\sqrt{{N_cg^2\over{3\pi}}} $  as the function of $N_c$ 
for the large values of $N_c$. 
In the 't Hooft model, the boson mass should be proportional to 
$\sqrt{N_cg^2\over{2\pi}}$, and therefore, the present expression of the boson mass 
is consistent with the 't Hooft evaluation 
as far as the expansion parameter is concerned. 
Therefore,  the boson mass calculation by the planar diagram evaluations 
of  't Hooft must be reasonable.   

Therefore, the boson mass prediction of 't Hooft should be reexamined 
from the point of view of  the light cone procedure.  It seems that the 
 't Hooft equations in the light cone have lost one important 
information which is expressed in terms of the $\theta_p$ variables 
both in the paper by Bars and Green \cite{q211} and also in the present paper. 
Since the variables $\theta_p$ are closely  related to the condensate 
values, the equations without the $\theta_p$ variables should correspond 
to the trivial vacuum in our point of view. Therefore, if one can recover 
this constraint in the 't Hooft equations in the light cone, 
then one may obtain the right boson mass from the  't Hooft model. 


\subsection{ RPA Calculations in  QED$_2$ and  QCD$_2$}

Up to this point, we have presented the calculated results of the Fock space 
expansion with the Bogoliubov vacuum state for  QED$_2$ and  QCD$_2$. 
The lowest  boson mass which is calculated by the Fock space expansion 
must be exact for the fermion and anti-fermion states 
if the vacuum is exact. From the present result for the condensate values 
of  QED$_2$ and  QCD$_2$, it indicates that the Bogoliubov vacuum state 
should be very good or may well be exact. 

On the other hand, there are  boson mass calculations by employing 
the Random Phase Approximation (RPA) method, and  
some people believe that the RPA calculation should be better than 
the Fock space expansion. 

Therefore, in this section, we present our calculated results of the RPA equations 
for  QED$_2$ and QCD$_2$ since there are no careful calculations in the 
very small fermion mass regions. First, we show that the RPA 
calculation for  QED$_2$ with the Bogoliubov vacuum state predicts the boson 
mass which is smaller than the Schwinger boson at the massless fermion limit. 
This means that the agreement achieved by the Fock space expansion is destroyed 
by the RPA calculation since it gives a fictitious attraction. 

Further, the RPA calculation for  QCD$_2$ with the Bogoliubov vacuum state 
produces an imaginary boson mass at the massless fermion limit. 
This is quite interesting, and it strongly suggests that the RPA equation 
cannot be reliable for fully relativistic cases 
since the eigenvalue equation of the RPA is not Hermitian,  
which is, in fact, a well known fact. 

Here, we briefly discuss the results of the RPA calculations, but  
the detailed discussion of the basic physical reason of the RPA problems 
will be given elsewhere. 

The RPA equations are based on the expectation that the backward moving 
effects of the fermion and anti-fermion may be included 
if one considers the following operator which contains 
the $d_{-m}c_m$ term in addition to the 
fermion and  anti-fermion creation term, 
$$ Q^\dagger = \sum_{n}(X_nc_n^\dagger d_{-n}^\dagger +Y_n d_{-n}c_n) . 
\eqno{(7.17)} $$
The RPA equations can be obtained by the following double commutations,
$$ \langle 0|[\delta Q,[H,Q^\dagger]]|0 \rangle = 
\omega \langle 0|[\delta Q,Q^\dagger]|0 \rangle  \eqno{(7.18)} $$
where $ \delta Q $ denotes 
$ \delta Q = d_{-n}c_n$ and $ c_n^\dagger d_{-n}^\dagger  $. 

Here, the vacuum $ |0 \rangle $ is assumed to satisfy the following condition, 
$$ Q |0 \rangle = 0 .  \eqno{(7.19)} $$
However, if the vacuum is constructed properly in the field theory model, 
it is impossible to find a vacuum that satisfies the condition of eq.(6.3). 
This fact leads to the RPA equations which are not Hermitian.

For QED$_2$, the RPA equations for $X_n$ and $Y_n$ become
\begin{align*}
{\cal M} X_n &= 2E_{n}X_n -\frac{g^2}{L}\sum_{m}X_{m}
       \frac{\cos^2(\theta_{n}-\theta_{m})}{(p_n-p_m)^2} \\
&-\lim_{\varepsilon\to 0}\frac{g^2}{L}\sum_{m}X_{m}
    \frac{\sin(\theta_{n-\varepsilon}-\theta_{n})
      \sin(\theta_{m}-\theta_{m-\varepsilon})}{\varepsilon^2} 
  -\frac{g^2}{L}\sum_{m}Y_{m}
  \frac{\sin^2(\theta_{n}-\theta_{m})}{(p_n-p_m)^2} \\
& -\lim_{\varepsilon\to 0}\frac{g^2}{L}\sum_{m}Y_{m}
 \frac{\sin(\theta_{n-\varepsilon}-\theta_{n})
      \sin(\theta_{m}-\theta_{m-\varepsilon})}{\varepsilon^2} \tag{7.20a} \\
-{\cal M} Y_n &= 2E_{n}Y_n -\frac{g^2}{L}\sum_{m}Y_{m}
    \frac{\cos^2(\theta_{n}-\theta_{m})}{(p_n-p_m)^2} \\
& -\lim_{\varepsilon\to 0}\frac{g^2}{L}\sum_{m}Y_{m}
   \frac{\sin(\theta_{n-\varepsilon}-\theta_{n})
      \sin(\theta_{m}-\theta_{m-\varepsilon})}{\varepsilon^2} 
  -\frac{g^2}{L}\sum_{m}X_{m}
   \frac{\sin^2(\theta_{n}-\theta_{m})}{(p_n-p_m)^2} \\
& -\lim_{\varepsilon\to 0}\frac{g^2}{L}\sum_{m}X_{m}
  \frac{\sin(\theta_{n-\varepsilon}-\theta_{n})
      \sin(\theta_{m}-\theta_{m-\varepsilon})}{\varepsilon^2} \tag{7.20b} 
\end{align*}
For QCD$_2$, one can easily derive the RPA equations, and 
at the large $N_c$ limit, they agree with the RPA equations 
which are obtained by Li et.al \cite{q03,q210}. 

It is important to note that the RPA equations 
are not Hermitian, and therefore there is no guarantee 
that the energy eigenvalues are real. In fact, as we  see below, 
the boson mass for QCD$_2$ becomes imaginary at the very small fermion mass. 


\begin{table}[h]
\caption{The masses for QED$_2$ and  QCD$_2$ with $SU(2)$ 
are measured by $ {g\over{\sqrt{\pi}}}$. 
The masses for large $N_c$ QCD$_2$ are measured by $ \sqrt{N_cg^2\over{2\pi}}$. 
$0.104 i$ indicates an imaginary eigenvalue.}

\begin{center}
\vspace{0.2cm}
\begin{tabular}{|c||c|c||c|c||c|c|}
\hline
\ & \multicolumn{2}{c||}{  QED$_2$} &
\multicolumn{2}{c||}{ QCD$_2$ SU(2)}  &
\multicolumn{2}{c|}{ Large $N_c$ QCD$_2$  \rule[-2mm]{0pt}{6mm} }  \\ 
\cline{2-7}
\ & $m_0=0$ &  $m_0 =0.1 $ &  $m_0=0$ &  $m_0 =0.1 $ 
&  $m_0=0$ &  $m_0 =0.1 $  \rule[-2mm]{0pt}{6mm} \\
\hline
\hline
\  & \  & \  & \  & \  & \  & \   \\
Fock Space & $1.000 $ &  $1.180$ &  0.467 &  $0.709$ 
&  0.543 &  $0.783$  \\
\  & \  & \  & \  & \ & \  & \   \\
\hline
\  & \  & \  & \  & \ & \  & \   \\
RPA & $ 0.989 $ 
&   $1.172$ 
&  $0.104 i $ & $ 0.576$ &  $0.120 i $ & $ 0.614$  \\
\  & \  & \  & \  & \ & \  & \   \\

\hline
\end{tabular} \\
\end{center}

\end{table}


In Table 5, we show the calculated values of the boson mass 
by the RPA equations for  QED$_2$ and  QCD$_2$ with the Bogoliubov vacuum state. 
It should be noted that the boson mass for $m_0=0$ case with the Fock space 
in the large $N_c$ limit is obtained from the 't Hooft equation. This equation is 
exactly the same as eq.(7.14) if we take the large $N_c$ limit. 
We note here that the boson mass (0.543  $ \sqrt{N_cg^2\over{2\pi}}$) 
at the large $N_c$ limit with the Fock space expansion just 
agrees with the value of eq.(7.2).

The behavior of the boson mass of the RPA calculation for  QCD$_2$ is not normal, 
contrary to the expectation. First, it is not linear as the function of $m_0$, but 
nonlinear in the small mass region. Further, the boson mass square becomes zero 
when the $m_0$ becomes a critical value, and it becomes negative 
when the $m_0$ is smaller than the critical value. In this case, 
the boson mass is imaginary, and thus this is physically not acceptable.  
This catastrophe is found to occur for the $SU(2)$ as well as 
for the large $N_c$ limit, as shown in Table 5.

At this point, we should comment on the belief that the RPA calculation 
should produce the massless boson at the massless fermion limit in QCD$_2$. 
However, if there were  physically a massless boson in two dimensions, 
this would be quite serious since a physical massless boson cannot propagate 
in two dimensions since it has an infra-red singularity in its propagator. 
But there is no way to remedy 
this infra-red catastrophe, and that is related to the theorem of  
Mermin, Wagner and Coleman \cite{q3,q4}. 
There are some arguments that the large $N_c$ limit is special because 
one takes the $N_c$ infinity. However, "infinity" in physics means simply 
that the $N_c$ must be sufficiently large, and in fact, as shown above, 
physical observables at $N_c=50$ are just the same as those of 
$N_c = \infty$. 
Therefore, it is rigorous that there should not exist any physical massless boson 
in two dimensions, even though one can write down the free massless boson Lagrangian 
density and study its mathematical structure. 
Thus, if one finds a massless boson constructed from the fermion and antifermion 
in two dimensions, then there must be something wrong in the calculations, 
and this is exactly what we see in the RPA calculations in  QCD$_2$.

In this respect, the boson mass calculated only by the Fock space expansion 
with the Bogoliubov vacuum can be reasonable from this point of view 
since there are some serious problems in the light cone as well as 
in the RPA calculations at the massless fermion limit. 

\subsection{  Spontaneous Chiral Symmetry Breaking in QCD$_2$}

The Lagrangian density of QCD$_2$ has a chiral symmetry 
when the fermion mass $m_0$ is set to zero. In this case, 
there should be no condensate for the vacuum state 
if the symmetry is preserved in the vacuum state. However, as we saw 
above, the physical vacuum state in QCD$_2$ has a finite condensate value, 
and thus the chiral symmetry is broken. In contrast to the Schwinger model, 
there is no anomaly in QCD$_2$, and therefore the chiral current is 
conserved. Thus, this symmetry breaking is spontaneous. 

However, there appears no massless boson.  Even though 
no appearance of the Goldstone boson is very reasonable 
in two dimension, this means that the Goldstone theorem does not 
hold for the fermion field theory. This is just what is proved in 
section 2 and 4, and the calculations of QCD$_2$ confirm its claim. 

Further, it seems that the chiral anomaly does not play an important 
role in the symmetry breaking business though it has been believed 
that the Schwinger model breaks the chiral symmetry due to the anomaly. 
However, the massless limit in QED$_2$ is not singular \cite{q11}. The condensate 
value and the boson mass are smooth as the function of the fermion mass $m_0$. 
This means that the vacuum structure is smoothly connected from the massive case 
to the massless one.  

This is just in contrast to the Thirring model \cite{q9,q12,q10} where 
the massless limit is a singular point as we saw in section 4. The structure of the vacuum is 
completely different from the massive case to the massless one 
in the Thirring model. 
Further, the condensate value and the boson 
mass in the Thirring model are not smooth function of the fermion mass $m_0$. 
For the massive Thirring model, there is no condensate, and the boson mass 
is proportional to  the fermion mass $m_0$ \cite{q13,q12,q10,q15}. 
Indeed, in the massive 
Thirring model, the induced mass term arising from the Bogoliubov transformation 
is completely absorbed into the mass renormalization term, and the vacuum 
stays as it is before the Bogoliubov transformation. 
But, for the massless Thirring model, the condensate is finite, 
and the condensate value and 
the boson mass are both proportional to the cutoff $\Lambda$ by which 
all of the physical observables are measured. 

On the other hand, QED$_2$ and QCD$_2$ are very different in that 
the coupling constant of the models have the mass scale dimensions, and 
all of the physical quantities are described by the coupling constant $g$ 
even at the massless limit. 
The super-renormalizability for QED$_2$ and QCD$_2$ must be quite important 
in this respect, while the Thirring model has no dimensional quantity, and this makes 
the vacuum structure very complicated when the fermion mass is zero. 

In Table 6, we summarize the physical quantities of the chiral symmetry breaking 
for QED$_2$, QCD$_2$ and Thirring models. All the condensates and the masses 
are measured in units of ${g\over{\sqrt{\pi}}}$ for QED$_2$ and QCD$_2$. 
The $\Lambda$ and $g_0$ 
in the Thirring model denote the cutoff parameter and the coupling constant, 
respectively. Also, the value of $ \alpha (g_0) $ can be obtained by 
solving the equation for bosons in the Thirring model \cite{q9,q10}.

For QED$_2$, there is an anomaly, and therefore, the chiral current 
is not conserved while, for QCD$_2$ and the Thirring model, the chiral 
current is conserved. From Table 6, one sees that the symmetry breaking 
mechanism is just the same for QED$_2$ and QCD$_2$. However, the Thirring 
model has a singularity at the massless fermion limit, and this gives 
rise to somewhat different behaviors from the gauge theory.


\begin{table}[htbp]
\caption{
We summarize the physical quantities of the chiral symmetry breaking 
for QED$_2$, QCD$_2$ and Thirring models.
}
\begin{center}
\vspace{0.2cm}
\footnotesize
\begin{tabular}{|c||c|c||c|c||c|}
\hline
\ & \multicolumn{2}{c||}{  Condensate} &
\multicolumn{2}{c||}{ Boson Mass  \rule[-2mm]{0pt}{6mm}} &
 Anomaly  \\ 
\cline{2-5}
\ & $m_0=0$ &  $m_0 \not= 0$ &  $m_0=0$ &  $m_0 \not= 0$ & \   \rule[-2mm]{0pt}{6mm}\\
\hline
\hline
\  & \  & \  & \  & \  & \    \\
QED$_2$ & $-0.283$ &  $-0.283+ O(m_0)$ &  1 &  $1+O(m_0)$ & yes  \\
\  & \  & \  & \  & \  & \    \\
\hline
\  & \  & \  & \  & \  & \    \\
QCD$_2$ & $-{N_c\over{\sqrt{12}}} \sqrt{N_c\over{2}}$ 
&   $-{N_c\over{\sqrt{12}}} \sqrt{N_c\over{2}}+O(m_0)$ & 
 ${2\over 3}\sqrt{{N_c\over{3}}}$ & $ \left( {2\over 3}{\sqrt{2\over 3}}
+{10\over{3}}{m_0\over{\sqrt{N_c}}} \right) 
\sqrt{{N_c\over{2}}}$ & no  \\
\  & \  & \  & \  & \  & \    \\
\hline
\  & \  & \  & \  & \  & \    \\
Thirring & ${\Lambda\over{g_0\sinh \left({\pi\over g_0}\right)}} $ 
& 0 &  ${\alpha (g_0)\Lambda\over{\sinh \left({\pi\over g_0}\right)}} $ 
 & $\alpha (g_0) m_0 $  & no  \\
\  & \  & \  & \  & \  & \    \\

\hline
\end{tabular}
\end{center}
\end{table}


\section{\bf Conclusions} 

The symmetry and its breaking in field theory were considered 
to be understood and settled down long time ago in terms of 
the Nambu-Goldstone theorem. In any of the field theory textbooks, 
the spontaneous symmetry breaking phenomena are well explained and described, 
and, therefore, it would have appeared somewhat odd to most of the readers 
to raise questions on the Goldstone boson after the spontaneous symmetry breaking. 

In this chapter, however, we have reviewed the recent progress in the spontaneous symmetry 
breaking and the appearance and non-appearance of a massless boson after 
the spontaneous symmetry breaking. Indeed, we have shown that the Nambu-Goldstone theorem 
for the fermion field theory is wrong. At the same time, we have presented examples 
of the symmetry breaking phenomena in concrete fermion field theory models which 
clearly exhibit the essence of the spontaneous symmetry breaking and 
the non-appearance of the massless boson associated with the symmetry breaking. 

The most important of the new aspects in the spontaneous symmetry breaking 
is to realize the change of the vacuum state after the symmetry breaking. 
This is of course quite well known since Nambu and Jona-Lasinio showed 
that the new vacuum is the one that breaks the chiral symmetry and 
its energy is lower than that of the symmetry preserving vacuum state (perturbative 
vacuum). In this case, however, one should be very careful for carrying out  
any field theory calculations since any physical estimations should be based 
on the physical vacuum state in quantum field theory. Here, it is for sure 
that one should start from the formalism that is based on the symmetry broken vacuum state 
since this is indeed a physical vacuum state. 

However, almost all of the calculations which were carried out for the NJL model 
are based on the perturbative vacuum state, and therefore the calculated 
results were not physical observables. This is rather a serious mistake, and 
in fact, people found a massless boson in the NJL model without noticing 
that their calculated boson is an unphysical particle. 
If one calculates the boson mass by the formulation which is 
based on the true vacuum ( symmetry broken vacuum ), then one finds 
that there is no massless boson. In fact, one obtains a massive boson 
if one carries out the calculation in terms of the Bogoliubov transformation method. 
In addition, we show that the Goldstone theorem cannot be applied to the fermion 
field theory models. This is obvious once we realize that the Goldstone theorem 
can only give the information on the dispersion relation of the energy and 
momentum for boson fields, and therefore it cannot tell us anything of 
the existence of the boson which should be constructed by fermions and antifermions. 
In fact, the Goldstone theorem had to assume the existence of the massless boson 
which is the one that should be obtained as the result of the proof. This is of course 
no proof at all for the fermion field theory models. Further, one notices that 
the commutation relation of the conserved charge and the boson field is obtained 
independently from the interaction Hamiltonian of the field theory model, and 
this is the basic ingredient of the proof of the Goldstone theorem. Therefore, 
it is clear that the commutation relation cannot give any information of 
the existence of the boson in the fermion field theory model. 

From these considerations, we can summarize the spontaneous symmetry breaking business 
in fermion field theory. The chiral symmetry of the fermion field theory 
models is spontaneously broken, and the chiral symmetry broken vacuum is indeed 
the physical state of the field theory model. This is all that happens in 
the chiral symmetry breaking. Even though Nambu and Jona-Lasinio claimed that 
the original massless fermion should become massive in the NJL model, 
the massless fermion should stay as it is. Under the Bogoliubov transformation method, 
it looks that the fermion should acquire the induced mass. But we believe 
that the fermion cannot change its structure by the spontaneous symmetry 
breaking phenomena and the massless fermions are still massless. 
The symmetry breaking is a property of the vacuum state, and 
the vacuum energy becomes lower than that of the symmetry preserving vacuum state. 
This energy change is entirely due to the change of the momentum distribution 
in the negative energy particles and it has nothing to do with the mass of the fermions 
in the vacuum state. Further, the renormalization procedure cannot give a finite 
mass to the massless fermions since the concept of the renormalization is just 
the change of the mass parameter from the infinite number to the finite observable.  

In the massless Thirring model, the spectrum with the symmetry preserving vacuum, 
though it is unphysical, has a gapless spectrum while the physical vacuum 
which breaks the chiral symmetry has the excitation spectrum with a finite gap. 
The gap in the excitation spectrum is due to the change in the vacuum structure. 
Even though this gap can be naturally explained by the massive fermions, 
it does not mean that the fermion becomes massive.  In fact, the Bethe ansatz 
solutions clearly show that the fermion stays massless even though the momentum 
distribution of the vacuum state is approximated by the dispersion of 
the free fermions to a good accuracy. 

If one employs the Bogoliubov transformation method in evaluating the boson 
in the massless Thirring model, then one finds a massive boson. 
However, the Bethe ansatz calculations show that there is no bosonic 
state in the massless Thirring model. Therefore, the massive boson which 
is predicted by the Bogoliubov transformation method in the NJL model 
should not exist in reality. This is not yet proved, but we believe 
that the claim must be quite reasonable since the Bogoliubov transformation 
method obviously overestimates the attraction between fermions and antifermions, 
and therefore they predict the bound state. In addition, the bound states are 
more difficult to make in four dimensions than in two dimensions, and this suggests 
that there should not be any bound states in the NJL model.

The nonexistence of bosons in the NJL and massless Thirring model should be 
closely related to the observation that the massless fermions and antifermions cannot 
make any bound states if the interactions are of the $\delta-$function type.   
This is in contrast to the gauge field theory models in two dimensions. 
For QED$_2$ and QCD$_2$, the massless fermions and antifermions make the bound 
states since they are confined. In other words, massless fermions can be either 
completely confined or cannot make any bound states since they do not have 
any rest systems.  

The physical connection between the chiral symmetry breaking and 
the chiral condensate is not clarified in this chapter. 
For the Thirring model, this is relatively simpler. The massive 
Thirring model has no chiral symmetry and there is no condensate. 
The vacuum state of the massive Thirring model is trivial. 
This can be seen at least from the Bogoliubov vacuum. 
Therefore, the massless limit is a singular point and the massless 
Thirring model has a chiral symmetry and its broken vacuum. 
In this chiral symmetry broken vacuum, the chiral condensate is finite. 
However, the chiral condensate value of the gauge field theory is finite 
even at the massive fermion case where there is no chiral symmetry. 
This suggests that the chiral symmetry and the chiral condensate is 
not strongly connected in the gauge field theory models.

In this chapter, we have not included the recent results on the Heisenberg $XXZ$ model. 
Therefore, we wish to make some comments on the relation between the massless 
Thirring model and the Heisenberg $XXZ$ model. It is believed that the two 
models are equivalent to each other since if one takes the massless limit in the 
Heisenberg $XYZ$ and massive Thirring models, then they can be reduced to the Heisenberg 
$XXZ$ model and the massless Thirring model, respectively. This belief is due to the fact  
that the equivalence between the Heisenberg $XYZ$ and massive Thirring models is well 
established \cite{lut}, and there is no problem over there.  

However, as we saw, the spectrum 
of the Thirring model gives a finite gap while the Heisenberg $XXZ$ model 
predicts always gapless spectrum. This means that, even though the two models 
are mathematically shown to be equivalent to each other, 
they are physically very different \cite{q92}. 

What should be the main reason for the difference ? 
If one makes the field theory into the lattice, then 
the lattice field theory loses some important continuous symmetry like Lorentz invariance 
or chiral symmetry. If the lost symmetry plays some important 
role for the spectrum of the model, then the lattice field theory becomes 
completely a different model from the continuous field theory model \cite{q98}. 
In general, the way of cutting the continuous space into a discrete one is not unique, 
and equal cutting of the space which is generally used in physics may not be 
sufficient for some of the field theory models. For example, if the interaction 
of the model is centered on the very small region of the space, then it is clear 
that the equal cutting of the space must be a very bad approximation. 

The non-equivalence between the Heisenberg $XXZ$ and massless Thirring models 
is the only example which shows a mismatch between the mathematical and physical 
correspondence. In this respect, we have discussed only a specific model 
in two dimensions, and it is quite different from four dimensional field theory models. 
However, the result certainly raises a warning on the lattice version 
of the field theory since clearly there are important continuous symmetries 
in any of the field theory models in four dimensions, and if these symmetries 
may be lost in the lattice version, then it is quite probable that the lattice 
calculations may not be able to reproduce a physically important spectrum 
of the continuous field theory models. 

Finally, we should like to discuss the symmetry breaking in four dimensional QCD. 
Since the real nature should be described by QCD in four dimensions, it is of course  
most important to understand what is the status of the symmetry breaking in QCD.  
First, let us assume that quarks are massless. In this case, there is the chiral 
symmetry, and we believe that the chiral symmetry should be broken in the vacuum. 
In other words, the energy of the symmetry broken vacuum state should be lower 
than the perturbative vacuum state which preserves the chiral symmetry. 
Since there is no scale in the four dimensional QCD, all the observables must be 
measured in terms of the cutoff momentum $\Lambda$. The vacuum structure is completely 
different from the perturbative one. However, we cannot say anything further than this. 
In a sense, the symmetry breaking phenomena in four dimensional QCD must be more similar 
to the massless Thirring model than QCD in two dimensions. But the dynamics is so 
complicated that we cannot build any realistic picture of the vacuum structure 
of the four dimensional QCD unless we solve the dynamics in a nonperturbative fashion. 

Now, in the realistic QCD in four dimensions, quarks have their own mass. 
In this case, there is no chiral symmetry, and therefore it does not make sense 
to argue the chiral symmetry breaking of the vacuum state. In this sense, 
QCD in four dimensions must have a singularity at the massless 
fermion limit, and the structure of the vacuum states between the massive fermion QCD 
and massless fermion QCD should be completely different from each other. In this respect, 
there is some similarity between QCD in four dimensions and Thirring model as far as 
the chiral symmetry breaking and its vacuum structure are concerned. But the structure 
of QCD in four dimensions must be much more complicated due to the gauge degree of freedom. 

In this sense, we have no idea about the structure of the vacuum in four 
dimensional QCD. We believe that the QCD vacuum must be quite complicated, but that is not 
related to the chiral symmetry breaking. All of the baryon and boson masses are measured 
in terms of the mass of the quarks. In this sense, the mass of pion (around 140 MeV) 
is quite large compared to the mass of $u$ or $d$ quarks (around 10 MeV) even though 
pion is lighter than other mesons by a factor of four or five.


\end{document}